\documentclass[journal=jpcafh,manuscript=article]{achemso}

\usepackage{longtable}
\usepackage{multirow}
\usepackage{amsmath}
\usepackage{subfigure}

\usepackage[usenames,dvipsnames]{color}
\usepackage{xcolor}
\usepackage{soul}
\usepackage{bm}
\usepackage{rotating}

\usepackage{amssymb}
\usepackage{titlecaps}
\Addlcwords{a an the of on at with for and in into from by without}
\usepackage{bm} 
\usepackage{outlines} 

\usepackage{xr}
\makeatletter

\newcommand*{\addFileDependency}[1]{
\typeout{(#1)}
\@addtofilelist{#1}
\IfFileExists{#1}{}{\typeout{No file #1.}}
}\makeatother

\newcommand*{\myexternaldocument}[1]{%
\externaldocument{#1}%
\addFileDependency{#1.tex}%
\addFileDependency{#1.aux}%
}

\myexternaldocument{si}

\SectionNumbersOn

\author{Meenu Upadhyay}

\author{Kai T\"opfer}

\author{Markus Meuwly}
\affiliation[University of Basel]{Department of Chemistry, University
  of Basel, Klingelbergstrasse 80, CH-4056 Basel, Switzerland.}
\email{m.meuwly@unibas.ch}

\title{Molecular Simulation for Atmospheric Reaction Exploration and
  Discovery: Non-Equilibrium Dynamics, Roaming and Glycolaldehyde
  Formation Following Photo-Induced Decomposition of {\it
    syn-}Acetaldehyde Oxide}

\begin{document}

\begin{abstract}
The decomposition and chemical dynamics for vibrationally excited {\it
  syn-}CH$_3$CHOO is followed based on statistically significant
numbers of molecular dynamics simulations. Using a neural
network-based reactive potential energy surface, transfer learned to
the CASPT2 level of theory, the final total kinetic energy release and
rotational state distributions of the OH fragment are in quantitative
agreement with experiment. In particular the widths of these
distributions are sensitive to the experimentally unknown strength of
the O--O bond strength, for which values $D_e \in [22,25]$ kcal/mol
are found. Due to the non-equilibrium nature of the process
considered, the energy-dependent rates do not depend appreciably on
the O--O scission energy. Roaming dynamics of the OH-photoproduct
leads to formation of glycolaldehyde on the picosecond time scale with
subsequent decomposition into CH$_2$OH+HCO. Atomistic simulations with
global reactive machine-learned energy functions provide a viable
route to quantitatively explore the chemistry and reaction dynamics
for atmospheric reactions.
\end{abstract}

\today

\section{Introduction}
Chemical processing and the evolution of molecular materials in the
atmosphere are primarily driven by photodissociation
reactions. Sunlight photo-excites the molecules in the different
layers of Earth's atmosphere and triggers chemical decomposition
reactions. The photoproducts are then reactants for downstream
reactions from which entire reaction networks
emerge.\cite{vereecken:2018}\\

\noindent
Within the extensive array of chemical reactions occurring in the
biosphere, Criegee intermediates (CIs) are one of the eminent reactive
species that have captured particular attention.\cite{rousso:2019}
Criegee intermediates are an important class of molecules generated in
the atmosphere from ozonolysis of alkenes which proceeds through a
1,3-cycloaddition of ozone across the C=C bond to form a primary
ozonide which then decomposes into an energized carbonyl oxide, also
known as CI, and one energized carbonyl (aldehyde or
ketone).\cite{criegee1949ozonisierung} The highly energized CIs
rapidly undergo either unimolecular decay to hydroxyl
radicals\cite{alam2011total} or collisional
stabilization\cite{novelli2014direct}. Stabilized CIs can isomerize
and decompose into products including the OH radical, or engage in
bimolecular reactions with water vapor, SO$_2$, NO$_2$ and
acids\cite{taatjes2017criegee,mauldin2012new}. In the laboratory,
generation of CIs in the gas phase has been possible from iodinated
precursors\cite{welz:2012,lester:2016} which allowed detailed
experimental studies of the photodissociation dynamics of {\it
  syn-}acetaldehyde oxide using laser
spectroscopy\cite{lester:2016,liu:2014,fang:2016} and provided
important information on its reactivity and decomposition dynamics,
including final state distributions of the OH
product. Computationally, a range of methods has been used, including
RRKM theory\cite{fang:2016} and MD simulations from the saddle point
separating {\it syn-}acetaldehyde oxide and vinyl hydroperoxide
(VHP).\cite{lester:2016} However, it was not until recently that the
entire reaction pathway from energized {\it syn-}CH$_3$COOH to OH
elimination was followed using neural network-based reactive potential
energy surfaces and molecular dynamics
simulations.\cite{MM.criegee:2021}\\

\begin{figure}[H]
\centering \includegraphics[scale=0.45]{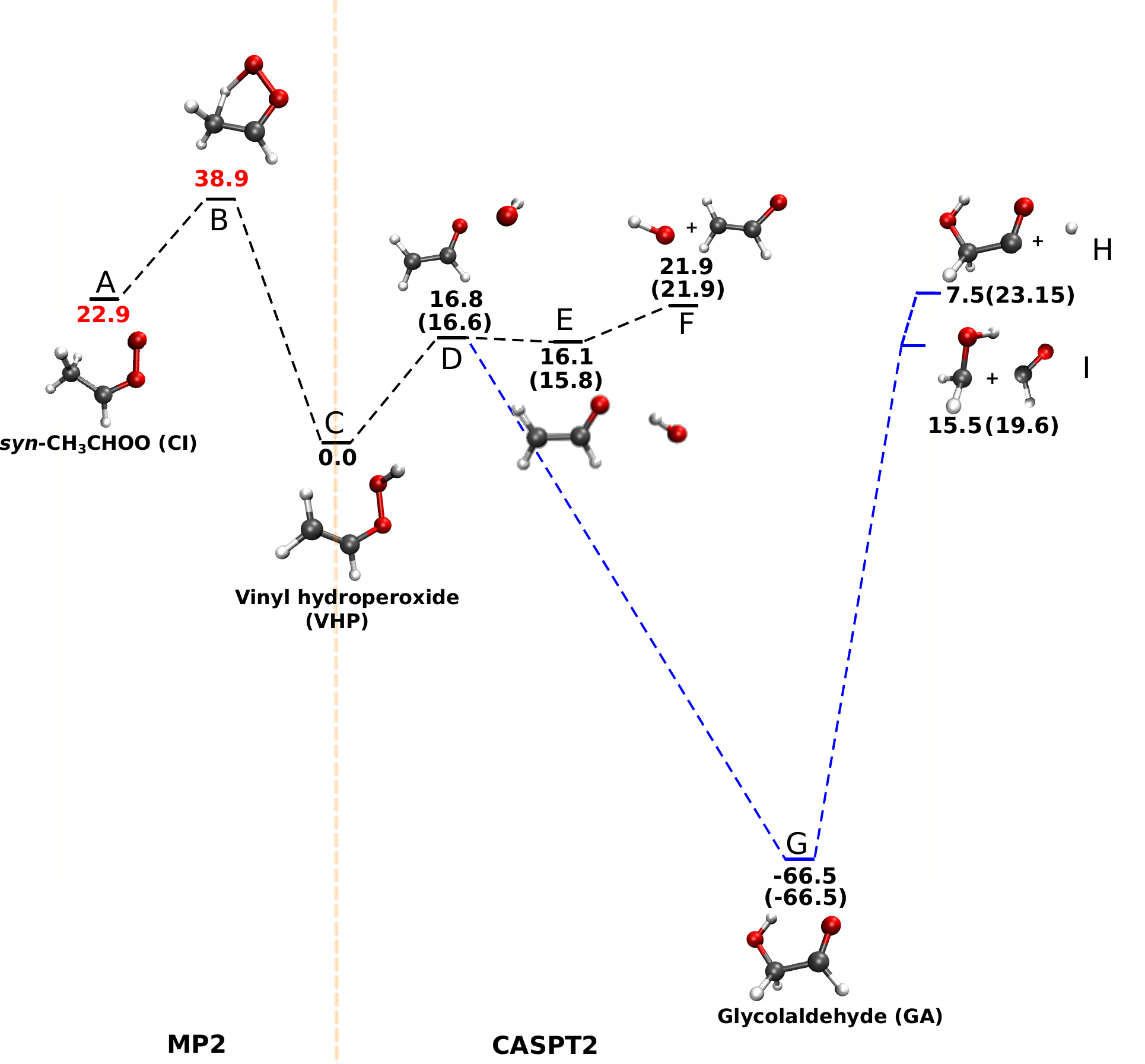}
    \caption{{\bf Schematic of potential energy surface relating
        species involved.} Energies (in kcal/mol) obtained at
      optimized geometries on PESs with VHP as the zero of energy. For
      the 1,4-hydrogen transfer, energies are reported from
      MP2/6-31G(d) level PES and upon VHP decomposition energies from
      CASPT2(12,10)/cc-pVDZ level PES are shown. The \textit{ab
        initio} reference energies are in brackets and those from the
      PhysNet PES are without brackets. The orange
      vertical line indicates the two models that were used in the
      hybrid simulations. For the decomposition of glycolaldehyde (GA) only 100
      structures for the CH$_2$OH+HCO channel were included in the
      data set whereas decay to CH$_2$OHCO+H was not covered by the
      model and trajectories leading to this channel were not
      analyzed.}
    \label{fig:ener_prof}
\end{figure}

\noindent
The ardent interest in understanding the photodissociation dynamics of
energized CIs, and in particular {\it syn-}CH$_3$CHOO, is the fact
that one of the decomposition products is the OH radical. The hydroxyl
radical, also referred to as the ``detergent of the
troposphere'',\cite{gligorovski2015environmental,levy1971normal} is
one of the most powerful oxidizing agents and plays an important role
in the chemical evolution of the atmosphere, triggering the
degradation of many pollutants including volatile organic compounds
(VOCs).\cite{stone:2012} Field studies have suggested that ozonolysis
of alkenes is responsible for the production of about one third of the
atmospheric OH during daytime, and is the predominant source of
hydroxyl radicals at night.\cite{emmerson2009night,khan2018criegee} In
addition to OH elimination, a second reaction pathway may lead to
glycolaldehyde (GA). Using molecular-beam mass spectrometry, the
formation of glycolaldehyde during the ozonolysis of trans-2-butene
was observed.\cite{conrad2021identification} This is the only
experimental work to our knowledge that reported glycolaldehyde
formation from the {\it syn-}CH$_3$CHOO Criegee
intermediate. Glycolaldehyde, an atmospheric volatile organic compound
can also be generated from isoprene\cite{lee2006gas},
ethene\cite{niki1981ftir} and biomass
burning\cite{bertschi2003trace}.\\

\noindent
The present work introduces atomistic simulations based on validated
machine-learned potential energy surfaces as a means for reaction
discovery for photodissociation processes. Such an approach includes
nonequilibrium, enthalpic, and entropic contributions to each of the
reaction channels included in the computational model.  For an
overview of all species considered in the present work with respective
relative energies, see Figure \ref{fig:ener_prof}. The combined use of
experimental and computed observables allows determination of the
elusive O--O scission energy for OH liberation following the initial
preparation of the reactant. For this, the final state total kinetic
energy release (TKER) and rotational distributions of the OH fragment
are determined from a statistically significant number of
full-dimensional, reactive molecular dynamics simulations, starting
from internally cold, vibrationally excited {\it syn-}CH$_3$CHOO with
two quanta in the CH stretch normal mode, akin to the experimental
procedures.\cite{lester:2016} The global potential energy surfaces are
either a neural network (NN) representation based on the PhysNet
architecture,\cite{MM.physnet:2019} or described at a more empirical
level, using multi state adiabatic reactive MD
(MS-ARMD).\cite{MM.armd:2014} Unlike MS-ARMD, with PESs based on
PhysNet chemical bonds can break and form akin to {\it ab initio} MD
simulations. This allows exploration of alternative reaction pathways
one of which is the formation of glycolaldehyde. The formation and
decomposition dynamics of GA is also studied in the present work and
expands on our view regarding the fate of activated ``simple''
molecules in the atmosphere.\\

\section{Results}

\subsection{Final State Distributions for the OH Fragment}
When considering computer-based tools for exploration of the dynamics
and chemical development of molecules, it is imperative to validate
the models used. Based on the MD simulations with the two PESs (see
Figure \ref{fig:ener_prof}) first the total kinetic energy release
(TKER) to CH$_2$CHO + OH products was analyzed. In trajectories
leading to OH as the final product, two different types of OH
elimination pathways were observed. The first is referred to as
``direct elimination'' whereas the second is classified as ``roaming
elimination''. A qualitative criterion based on the distance traveled
by the OH fragment was used to distinguish the two types of
processes. For this, the moment at which the O--O separation reaches 3
\AA\/ is set as the zero of time. Then, the time for elimination
($t_e$) required for the O--O distance to increase from 3 \AA\/ to 10
\AA\/ was recorded. For $t_{\rm e} < 0.8$ ps a trajectory was
identified as ``direct elimination'' whereas for all other cases
($t_{\rm e} \geq 0.8$ ps) it is a ``roaming elimination'', see SI and
Figures S2 to S4 for motivating this
choice for $t_{\rm e}$.\\

\begin{figure}
    \centering
    \includegraphics[scale=0.32]{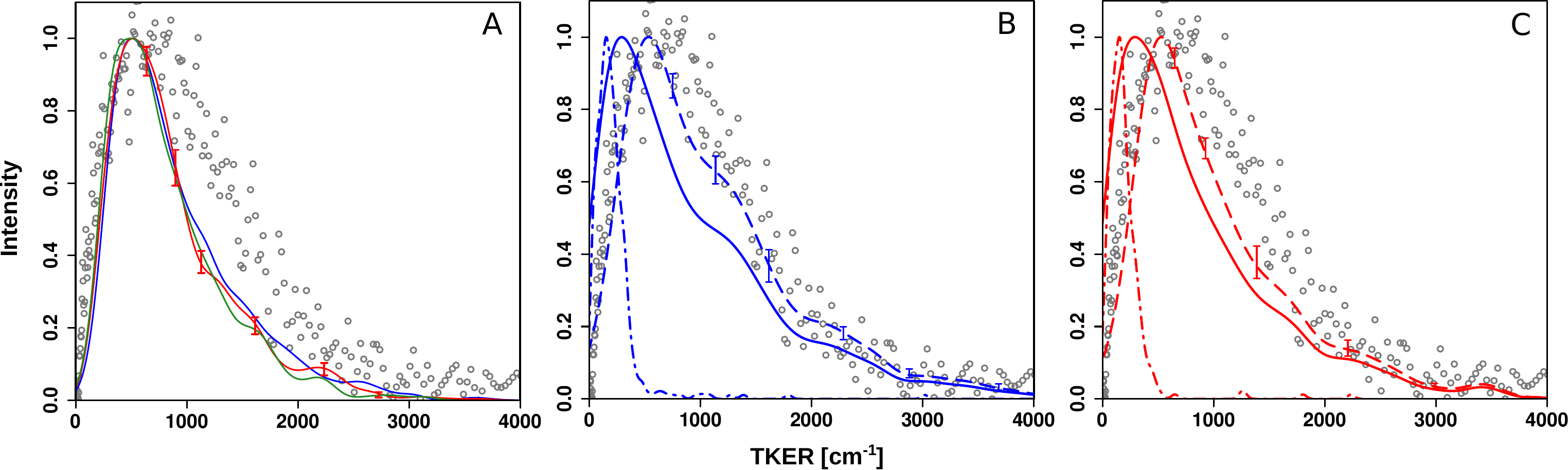}
   \caption{\textbf{TKER distribution to products from simulations and
       experiment.} The computed distributions (at 5988 cm$^{-1}$)
     forming OH(X$^2 \Pi$, $v=0$, all $N$) using different PESs
     compared with experimentally reported (grey
     circles)\cite{lester:2016} $P({\rm TKER})$ (at 6081 cm$^{-1}$)
     forming OH(X$^2 \Pi$, $v=0$, $N=3$). Panel A: MS-ARMD, Panel B:
     PhysNet with conventional $D_e = 22$ kcal/mol, Panel C: PhysNet
     with $D_e = 25$ kcal/mol. Solid, dotted-dashed and dashed lines
     correspond to all (direct+roaming), roaming and direct
     dissociated trajectories. Results are shown for $D_e = 22$
     kcal/mol (blue), $D_e = 25$ kcal/mol (red), and $D_e = 27$
     kcal/mol (green). For MS-ARMD all OH formed through direct
     dissociation. Computed error bars are from bootstrapping and the
     maximum of the distribution is set to unity. For dissociation to
     OH(X$^2 \Pi$, $v=0$, $N=3$) see Figure S5.}
    \label{fig:tker}
\end{figure}

\noindent
Figure \ref{fig:tker} compares TKER distributions from simulations
with CH-excitation at 5988 cm$^{-1}$ (because experimentally $P(N)$
was reported at this excitation energy, see below) using different
PESs with experimentally determined $P({\rm TKER})$ at a CH-excitation
energy of 6081 cm$^{-1}$ forming OH(X$^2 \Pi$, $v=0$, $N=3$). For
reference, from a 6-point moving average the position of the
experimentally measured maximum is found at 542 cm$^{-1}$.  Figure
\ref{fig:tker}A reports distributions obtained from using the MS-ARMD
PES for $D_e = 22$, 25, 27 kcal/mol (blue, red, green traces). The
computed $P({\rm TKER})$ qualitatively correctly capture the
experimentally determined distribution but are somewhat narrower with
a peak at $\sim 500$ cm$^{-1}$ shifted to lower energy by $\sim 40$
cm$^{-1}$ compared with experiments. Additional exploratory
simulations were also carried out for a CH-excitation energy of 6081
cm$^{-1}$ using MS-ARMD and no significant change in the peak position
was observed.\\

\noindent
On the other hand, $P({\rm TKER})$ to yield OH(X$^2 \Pi$, $v=0$, all
$N$) for ``direct elimination'' from simulations using the PhysNet PES
with two different dissociation energies for O--O scission agree well
with experiment, see Figures \ref{fig:tker}B and C. the agreement is
even better if only the subpopulation forming OH(X$^2 \Pi$, $v=0$,
$N=3$) is compared with experiment, see Figure S5. In
Figures \ref{fig:tker}B ($D_e = 22$ kcal/mol) and \ref{fig:tker}C
($D_e = 25$ kcal/mol) reports $P({\rm TKER})$ from ``direct
elimination'' (dashed), ``roaming elimination'' (dotted-dashed) and
their sum (solid). With $D_e = 22$ kcal/mol the peak position for
``direct elimination'' is at 538 cm$^{-1}$ - in almost quantitative
agreement with experiment - which shifts to 290 cm$^{-1}$ if all
trajectories are analyzed. Most notably, the total width of $P({\rm
  TKER})$ realistically captures the distribution measured from
experiments, including undulations at higher energies. For an O--O
scission energy of $D_e = 25$ kcal/mol the maximum for directly
dissociating trajectories is at 531 cm$^{-1}$, compared with 286
cm$^{-1}$ if all trajectories are analyzed. The small shift of
$\approx$ 5 cm$^{-1}$ in the peak position on increasing the barrier
by 3 kcal/mol shows that the TKER does not depend appreciably on the
barrier for O-O dissociation. This is because independent of the
precise value of $D_e$, the van der Waals complex in the product
channel (CH$_2$CHO---HO) is always stabilized by 5 kcal/mol
(structures E and F in Figure \ref{fig:ener_prof}) relative to the
completely separated CH$_2$CHO + OH fragments. On the other hand, the
{\it width} of $P({\rm TKER})$ responds directly to changes in $D_e$ -
with increasing dissociation energy, the width of the distribution
decreases appreciably. The observation that the sum of ``direct
elimination'' and ``roaming trajectories'' yields a peak shifted to
lower energy and less favourable agreement with experiment indicates
that the amount of roaming trajectories ($\sim 40$ \%) may be
overestimated in the present simulations. Recent experiments and
calculations on the formation of 1-hydroxy-2-butanone from {\it
  anti-}methyl-ethyl substituted Criegee intermediate found $\leq 10$
\% contribution of a roaming pathway.\cite{lester:2023} This
overestimation may be due limitations in the CASPT2/cc-pVDZ
calculations to accurately describe long range interactions to yield
an overproportion of ``roaming trajectories''.\\

\noindent
Compared to simulations using the PhysNet PES, the MS-ARMD PES allows
qualitatively correct studies but can, e.g., not describe the
``roaming" part of the dynamics. Furthermore, quantitative studies
require more detailed information about the angular/radial coupling
and the electrostatic interactions as afforded by the PhysNet model
based on CASPT2 calculations. Another distinguishing feature between
the MS-ARMD and PhysNet PESs is that the NN-trained model features
fluctuating partial charges depending on the internal degrees of
freedom of the species involved. From a dynamics perspective,
trajectories based on PhysNet convincingly demonstrate that
OH-elimination proceeds either through a ``direct" or a
``roaming-type" mechanism. Hence, the added effort in constructing the
PhysNet PES and the considerably increased compute time incurred
compared with the computationally efficient MS-ARMD simulations allow
to move from qualitative to quantitative results, both in terms of
$P({\rm TKER})$ and the mechanistic underpinnings derived from
analysis of the trajectories.\\

\begin{figure}
    \centering
    \includegraphics[scale=0.42]{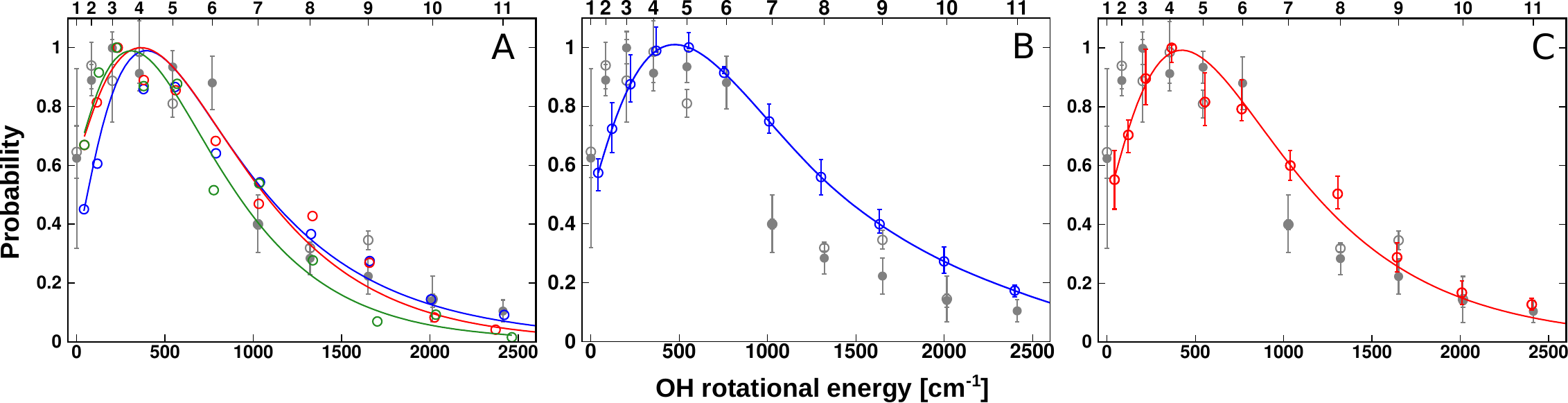}
    \caption{\textbf{OH product rotational energy and state
        distribution from simulations and experiment.} Rotational
      energy and corresponding state distribution at for CH excitation
      energy of 5988 cm$^{-1}$ compared with experiments for OH
      radicals. The solid grey circles with error bars are from the
      experiments. Panel A: MS-ARMD, Panel B: PhysNet with
      conventional $D_e = 22$ kcal/mol, Panel C: PhysNet with $D_e =
      25$ kcal/mol. Here solid line corresponds to all (roaming +
      direct) dissociated trajectories. Color code: blue = $D_e = 22$,
      red = $D_e = 25$ and green = $D_e = 27$ kcal/mol. Error bars
      from bootstrapping are included and the most probable state is
      set to unity.}
    \label{fig:comberot}
\end{figure}

\noindent
Final OH-rotational distributions following CH-excitation at 5988
cm$^{-1}$ are shown in Figures \ref{fig:comberot}A (MS-ARMD) and B/C
(PhysNet with different values for $D_e$). The experimental $P(N)$
distribution (grey symbols), also reported from excitation with 5988
cm$^{-1}$, peaks at $N = 3$ although the peak is broad and may also be
considered to extend across $3 \leq N \leq 6$. This compares with
$N_{\rm max} = 4$ from the simulations on the MS-ARMD PESs with $D_e =
22$, 25, and 27 kcal/mol (solid blue, red, green circles), see Figure
\ref{fig:comberot}A. The computed $P(N)$ follow a gamma-distribution
(solid lines) which represent waiting time
distributions\cite{olguin:2021} and were successfully used to model
photodissociation of H$_2$SO$_4$ following vibrational excitation of
the OH-stretch.\cite{yosa:2011,reyes.pccp.2014.msarmd} With increasing
$D_e$ the $N_{\rm max}$ shifts to lower values: for $D_e = 22$
kcal/mol $P(N)$ peaks at $N= 4$ and decreases to $N= 3$ for $D_e = 25$
and 27 kcal/mol. For all trajectories that were analyzed, the final
vibrational energy of the OH product was always close to its
vibrational ground state, i.e. $\nu_{\rm OH} = 0$. \\

\noindent
Results from simulations using PhysNet with $D_e = 22$ and 25 kcal/mol
in Figures \ref{fig:comberot}B and C report $P(N)$ for OH. Both
distributions realistically describe the experimentally reported
$P(N)$, in particular the plateau between $3 \leq N \leq 6$ - and for
$D_e = 25$ kcal/mol also the width is captured. For $D_e = 22$
kcal/mol the width of $P(N)$ is somewhat larger than that from
experiments. With increasing dissociation energy the maximum for
$P(N)$ shifts from $N_{\rm max} = 5$ to $N_{\rm max} = 4$. The $P(N)$
for ``direct" and ``roaming" type dissociation are shown in Figure
S6. The final $N$-state of the photodissociating OH-product
is to some extent correlated with the average O-O-H angle at the
moment of dissociation which is assumed to be at an O--O separation of
3 \AA\/. For final OH-rotational states $N < 6$ the average O-O-H
angle decreases from $140^\circ$ to $105^\circ$ whereas for $N \geq 6$
it remains at $\sim 100^\circ$, see Figure S7.\\

\noindent
Previous computational studies\cite{lester:2016} reported final state
distributions from trajectories that were initiated from either the
H-transfer TS or the submerged saddle point configuration immediately
preceding the OH elimination step D, see Figure \ref{fig:ener_prof},
i.e. from simulations that did not follow the entire dynamics between
{\it syn-}CH$_3$COOH to OH-elimination. Simulations starting at the
H-transfer TS yielded total kinetic energy release distributions
$P({\rm TKER})$ for OH(X$^2 \Pi$, $v=0$, all $N$) with the maximum
peak shifted to considerably higher energy ($\sim 1500$ cm$^{-1}$
compared with $\sim 540$ cm$^{-1}$ from experiments) and a width
larger by a factor of $\sim 2$ compared with those observed, whereas
starting trajectories at the submerged barrier shifted the maximum to
lower kinetic energy ($\sim 800$ cm$^{-1}$). For the OH-rotational
distributions $P(N)$, it was found that starting the dynamics at the
H-transfer transition state yields a width and position of the maximum
in accordance with experiments whereas initiating the dynamics at the
submerged saddle point shifts the maximum to smaller $N_{\rm max}$ and
leads to narrower $P(N)$. Taken together, starting the dynamics at
intermediate yields qualitatively correct final state distributions
which can, however, differ by up to a factor of 3 for position of the
peak maximum or width from the experimentally measured ones.\\

\noindent
Taken together, the present results for $P({\rm TKER})$ and $P(N)$
suggest that PhysNet transfer-learned to the CASPT2 level and used in
the hybrid simulations with $D_e = 22$ kcal/mol for the O--O bond
yields final state distributions consistent with experiment. With
increasing $D_e$ from 22 to 25 kcal/mol both final state distributions
show decreasing probabilities for higher kinetic energies and
rotational states when considering OH(X$^2 \Pi$, $v=0$, all $N$). If
the final rotational state of the OH-photoproduct is limited to $N=3$
and compared with experiment (see Figure S5), both
values of $D_e$ yield favourable agreement with experiment, in
particular for ``directly dissociating'' trajectories, whereas for
$D_e = 25$ kcal/mol $P({\rm TKER})$ from ``all trajectories'' (solid
red line) yields good agreement as well. Furthermore, the simulations
provide information about the $N-$dependence of $P({\rm TKER})$, see
Figure S8, which indicates that for OH$(v=0, N
\in [0,3])$ the distribution is somewhat more narrow up to ${\rm TKER}
\sim 1000$ cm$^{-1}$ with more pronounced undulations compared with
$P({\rm TKER})$ for OH($v=0$, $N \in [9,12]$). This is information
complementary to what is available from experiments. The MS-ARMD
simulations yield $P({\rm TKER})$ that are somewhat too narrow but
$P(N)$ with $D_e = 25$ kcal/mol also realistically describes the
experimentally reported distributions. These findings support the
present and previous CASPT2 calculations that found and reported
dissociation energies of up to 26 kcal/mol and clarify that 31.5
(6-31G(d)) and 35.7 (aug-cc-pVTZ) kcal/mol from MP2 calculations
suffer from assuming a single-reference character of the electronic
wavefunction.\cite{lester:2016} \\

\subsection{Thermal Rates}
With the new PhysNet PES transfer-learned to the CASPT2 level,
energy-dependent rates were determined for 3 excitation energies
i.e. [5603, 5818, 5988] cm$^{-1}$. The stretched exponential with
stretch exponents ranging from 1.4 to 1.9 remarkably well captures the
behavior of $N(t)$. Such values suggest that kinetic intermediates are
visited that lead to distributed
barriers,\cite{austin:1975,MM.mbno:2016} which is also consistent with
the shape of $P(N)$ which are Gamma-distributions.\\

\begin{figure}
\centering \includegraphics[scale=0.51]{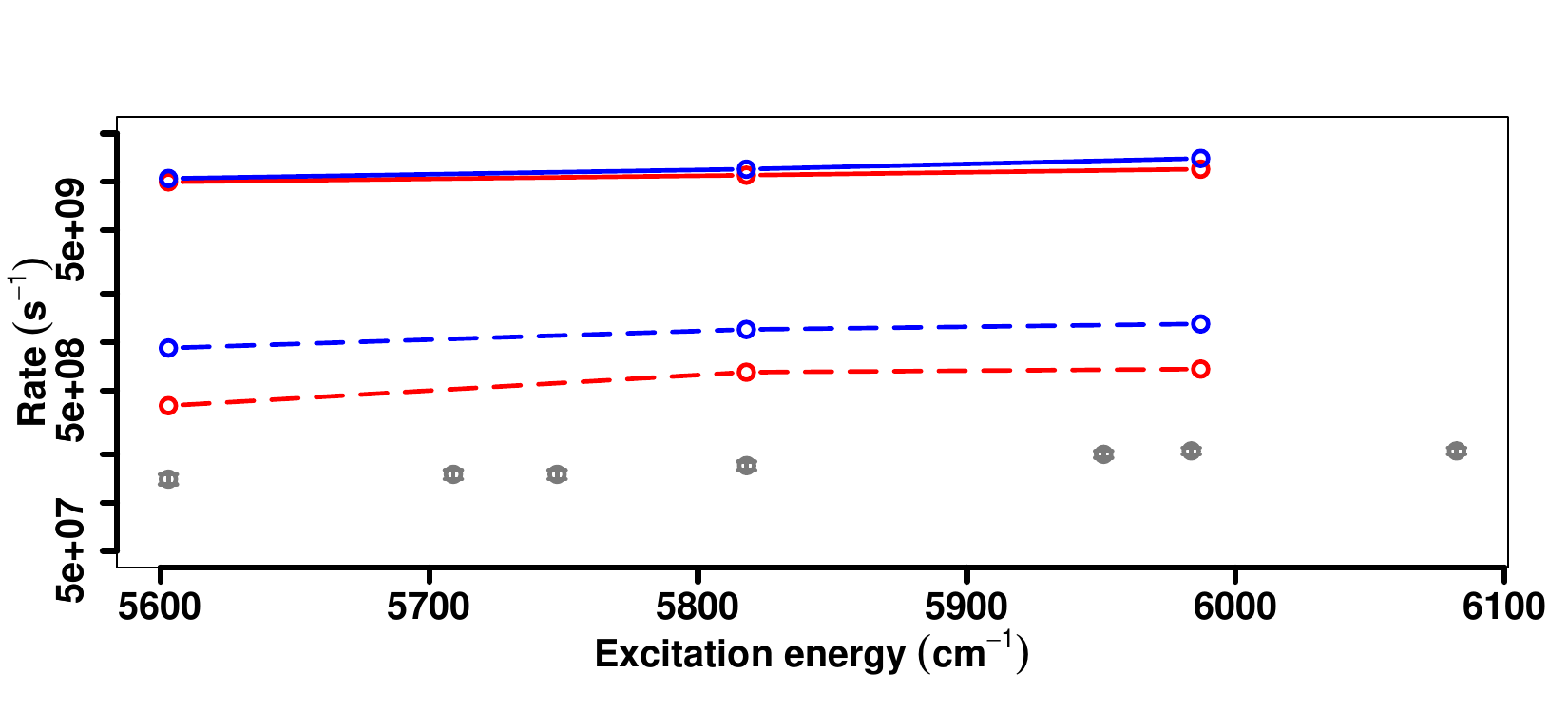}
   \caption{\textbf{OH formation rates using PhysNet PES compared with
       the experiments.} The grey points with error bars are the
     experimental results (Table S1).\cite{fang:2016} The red and blue
     symbols are for $D_e = 22$ and 25 kcal/mol, respectively. Dashed
     and solid lines are rates based on fitting $N(t)$ to single and
     stretched exponential decays, respectively, see Figure
     S9. Experimental rates are 1.4 to 2.1$\times 10^8$
     s$^{-1}$ for excitation energies between 5603 and 6082 cm$^{-1}$,
     compared with 4.0 to $6.8 \times 10^8$ with $D_e = 25$ kcal/mol
     and $9.2 \times 10^8$ to $1.3 \times 10^9$ with $D_e = 22$
     kcal/mol using single exponential function.}
    \label{fig:rate}
\end{figure}

\noindent
The thermal rates for the two values of $D_e$ and from using the two
fitting functions are reported in Figure \ref{fig:rate} together with
the experimentally determined rates. The moderate energy dependence of
$k(E)$ is correctly captured by the simulations but - depending on
whether a single- or a stretched-exponential is employed - the
absolute values of the rates are too large by one to two orders of
magnitude, respectively. Within equilibrium transition state theory
this reflects a barrier along the pathway that is overestimated by 2
to 3 kcal/mol. This point is discussed further below.\\

\subsection{Dissociating Trajectories, Roaming and Glycolaldehyde Formation}
\begin{figure}
\centering \includegraphics[scale=0.55]{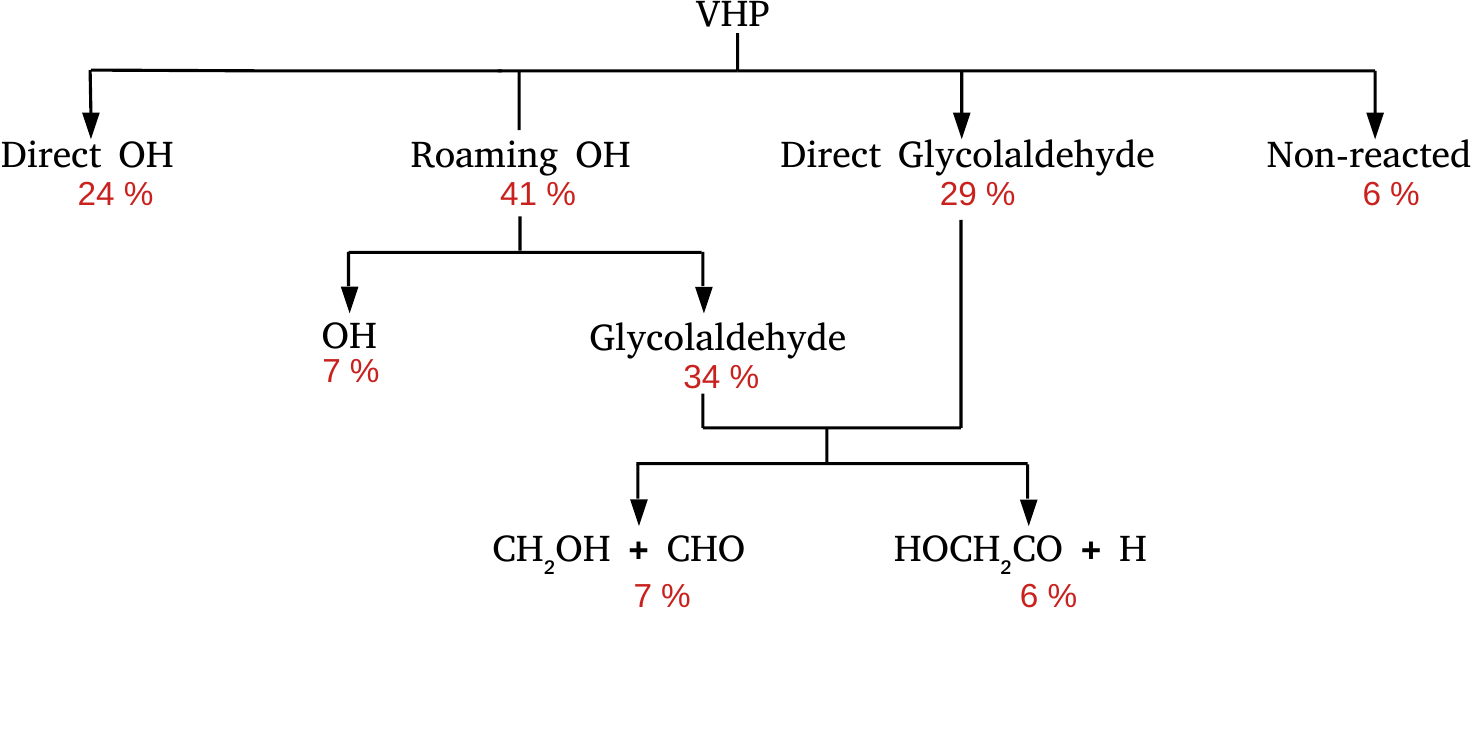}
\caption{\textbf{Branching fraction of the final products formed using
    PhysNet PES from simulations on the 200 ps time scale.} Branching
  ratio of CH$_2$CHO, OH, CH$_2$OH, HCO radicals and glycolaldehyde
  products observed from VHP using PhysNet PES with $D_e = 22$
  kcal/mol at 5988 cm$^{-1}$. OH radical is formed from both direct
  and roaming trajectories and HCO + CH$_2$OH is observed from
  glycolaldehyde.}
\label{fig:pathways}
\end{figure}

Following the dynamics of OH after dissociation revealed that the
system accesses final states different from CH$_2$CHO+OH. As one
alternative, roaming of OH leads to recombination with the CH$_2$
group to yield glycolaldehyde which is stabilized by 66.5 kcal/mol
relative to VHP, see Figure \ref{fig:ener_prof}. This intermediate can
decay to form new products such as CH$_2$OH+HCO or CH$_2$OHCO+H, see
Figures \ref{fig:ener_prof} and \ref{fig:pathways}.\\

\noindent
An overview of the decay channels observed from the simulations using
the PhysNet PES following formation of internally hot VHP is shown in
Figure \ref{fig:pathways}. It is useful to recall that the reaction is
initiated through excitation of the CH-stretch mode with approximately
2 quanta which makes elimination of OH from VHP a nonequilibrium
process because vibrational energy redistribution is not complete on
the lifetime of VHP ($\sim 20$ ps, see Figure
S10). For each pathway and product the percentages of
the final states on the 200 ps time scale are reported which were
determined from 6000 independent trajectories with $D_e = 22$ kcal/mol
and excitation of the CH-stretch at 5988 cm$^{-1}$. Out of all the
trajectories run (excluding those with OH-vibrational energies lower
than the quantum-mechanical zero-point energy), 24 \% lead to direct
elimination of OH and 41 \% roam around CH$_2$CHO out of which 20 \%
(i.e. 7 \% of the total) lead to OH elimination and 80 \% (i.e. 34 \%
of total) form GA. The remaining 29 \% yield GA in a direct fashion
and 6 \% remain in VHP on the 200 ps time scale but may undergo
further chemical processing on longer time scales. From the total GA
population formed (63 \% of all trajectories), 7 \% decay to CH$_2$OH
+ HCO, 6 \% end up in CH$_2$OHCO+H and the remainder in GA. It should
be noted that branching fractions depend on the total simulation time
considered. For example, it is expected that all of GA formed in the
gas phase will eventually decay on sufficiently long time scales
because sufficient energy is available.\\

\begin{figure}
\centering
    \includegraphics[scale=0.6]{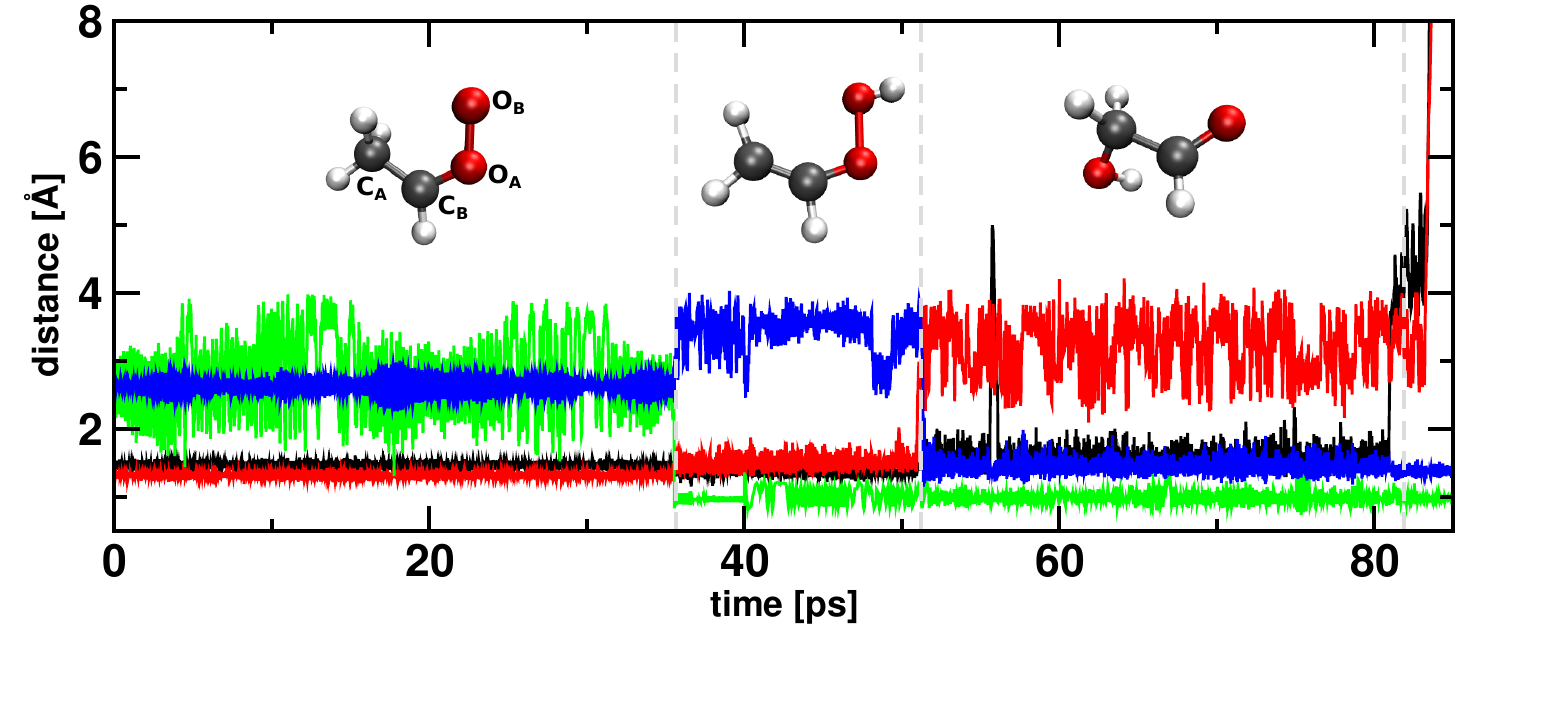}
    \caption{\textbf{Time series for different atom-atom separations
        for the reaction \textit{syn}-CH$_3$CHOO $\longrightarrow$
        HCO+CH$_2$OH.} Time series for C$_{\rm A}$-C$_{\rm B}$
      (black), O$_{\rm A}$-O$_{\rm B}$ (red), O$_{\rm B}$-H (green)
      and C$_{\rm A}$-O$_{\rm B}$ (blue) separations for a trajectory
      leading to HCO+CH$_2$OH radical formation using PhysNet PES with
      $D_e = 22$ kcal/mol at 5988 cm$^{-1}$. At 36 ps VHP is formed
      followed by GA formation and equilibration for the next 30 ps
      and at 81 ps C-C bond dissociates to form HCO+CH$_2$OH.}
    \label{fig:timeseries-hco}
\end{figure}

\noindent
Explicit time series for important bonds involved in formation of GA
are reported in Figure \ref{fig:timeseries-hco}. Starting from {\it
  syn}-CH$_3$COOH at $(t=0)$, vibrational excitation of the CH-stretch
leads to formation of VHP after 36 ps. The system dwells in this state
for $\sim 15$ ps until OH elimination and transfer to the CH$_2$ group
leads to GA with a residence time of $\sim 30$ ps by which HCO
elimination occurs at 81 ps. The trajectory shown involves direct
OH-transfer between VHP and GA, without OH-roaming.\\

\begin{figure}
\includegraphics[scale=0.40]{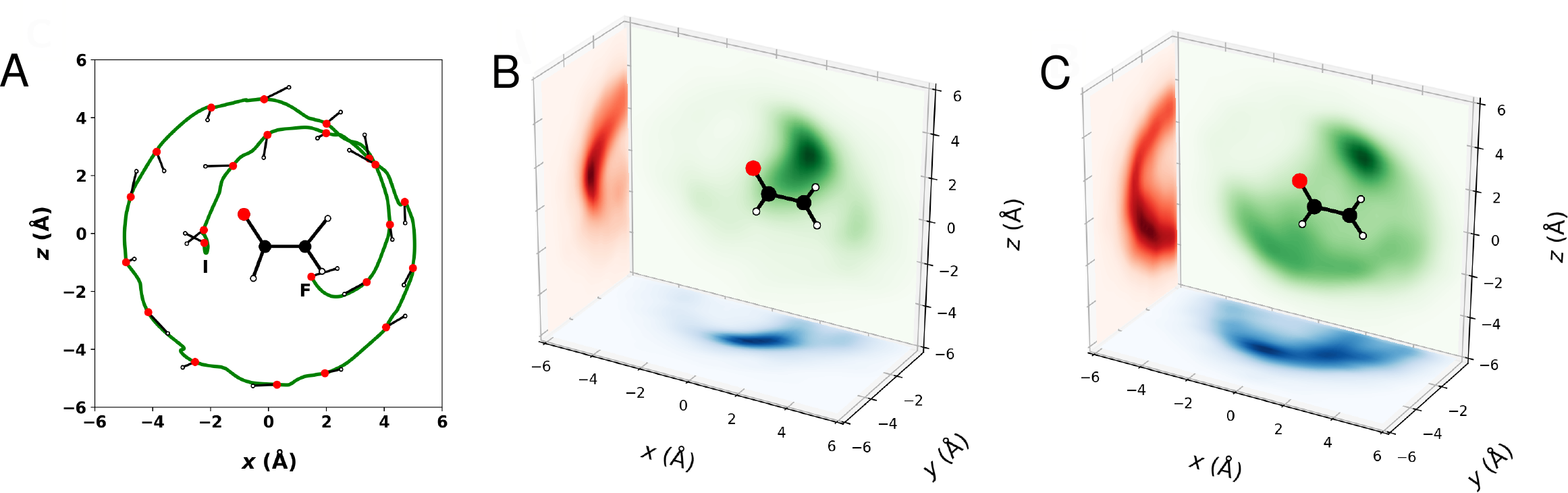}
    \caption{\textbf{OH roaming around vinoxy before glycolaldehyde
        formation.} Panel A: OH roaming trajectory projected on X-Z
      plane around CH$_{2}$CHO before glycolaldehyde formation using
      PhysNet PES with $D_e = 22$ kcal/mol at 5988 cm$^{-1}$ for 1
      ps. Here, \textbf{I} denotes the initial position of OH at which
      O-O bond of VHP dissociates and \textbf{F} denotes the final
      position of OH before glycolaldehyde formation. Panels B and C:
      Probability densities of the OH radical around the CH$_{2}$CHO
      group (with C$_{\rm A}$-C$_{\rm B}$-O$_{\rm A}$ in the $xz$
      plane and the center of mass of CH$_{2}$CHO in the origin) from
      $\sim 500$ trajectories leading to B: direct or C: roaming
      glycolaldehyde formation on the PhysNet PES with $D_e = 22$
      kcal/mol. Trajectories in panel C access the half planes
      $(-x,y)$ and $(-x,z)$ which are not visited in panel B for
      ``direct transfer" trajectories.}
    \label{fig:roam}
\end{figure}

\noindent
Figure \ref{fig:roam}A shows an example of OH dynamics around the
vinoxy radical before glycolaldehyde formation. This ``attack from the
backside'' is a hallmark of a roaming reaction. Probability densities
for the OH radical moving around CH$_2$CHO before glycolaldehyde
formation using 500 independent trajectories from PhysNet PES with
$D_e = 22$ kcal/mol are shown in Figures \ref{fig:roam}B/C. In both
``direct transfer" (Figure \ref{fig:roam}B) and ``roaming" (Figure
\ref{fig:roam}C) trajectories, the OH radical moves out-of-plane to
attack the -CH$_2$ group. A similar out-of-plane roaming was reported
from experiments on formaldehyde.\cite{foley2021orbiting} In direct
transfer trajectories, OH follows a semicircular path around the
vinoxy group whereas pronounced roaming is followed for the other
trajectories, see Figure \ref{fig:roam}C.\\

\noindent
Formation times of GA after dissociation of OH from VHP extend up to
$\sim 2$ ps, see Figure S11. The initial,
pronounced peak at 0.1 ps is due to ``direct transfer'' between VHP
and GA whereas the remainder of the exponentially decaying
distribution involves OH-roaming trajectories with maximum diffusion
times of 2.6 ps. Out of the total amount of GA formed, 11 \% (7 \% of
total VHP) show C-C bond cleavage to form CH$_2$OH + HCO radicals and
the lifetime distribution of GA before dissociation is shown in Figure
S12. The distribution has lifetimes of up to $\sim 10$
ps with a most probable lifetime around $\sim 1$ ps.  Two-dimensional
projections of the OH positions around the CH$_2$CHO radical as shown
in Figure \ref{fig:roam} provide an impression of the spatial range
sampled by the OH radical before formation of GA. Typical excursions
involve separations of 3 to 4 \AA\/ away from the center of the C-C
bond of CH$_2$CHO. \\

\section{Discussion and Conclusion}
The present work establishes that photodissociation of the energized
{\it syn-}acetaldehyde oxide intermediate can lead to additional
products beyond the known CH$_2$CHO+OH fragments. This is possible
because the OH fragment following dissociation of VHP can roam around
the CH$_2$CHO radical and lead to glycolaldehyde formation from which
other dissociation channels open up, e.g. leading to the CH$_2$OH+HCO
or CH$_2$OHCO+H fragments. This discovery was possible because
NN-trained PESs allow bond-breaking/bond-formation to occur akin to
{\it ab initio} MD simulations but allowing for exploration of
statistically significant numbers of simulations to be run.\\

\noindent
Two strategies were used in the present work to refine and improve the
PESs\cite{MM.criegee:2021}. The first one was based on adjusting the
dissociation energy for OH formation in the MS-ARMD PES, akin to
``morphing'',\cite{MM99:nehf,bowman:1991} which is particularly simple
due to the empirical nature of the parametrization. The second
approach was to use multi-reference-based training energies (at the
CASPT2 level) to transfer learn the reactive PES based on the PhysNet
architecture. A multi-reference treatment is required to realistically
describe the energetics for VHP, its dissociation to CH$_2$CHO+OH and
reformation and decay of GA. Both approaches adapt the shape of a
global PES based on local information and attempt to preserve the
overall shape of a PES by fine-tuning features to which the available
observables are sensitive.\cite{MM99:nehf,bowman:1991}\\

\noindent
The product state distributions for fragment kinetic energy and
rotational excitation of the OH product are in almost quantitative
agreement with experiment, in particular for $D_e = 22$ kcal/mol and
the transfer-learned PhysNet PES. This differs from earlier
efforts\cite{lester:2016} which initiated the dynamics at transition
states and found only qualitative agreement, in particular for $P({\rm
  TKER})$. Even the more empirical, MS-ARMD representation yields
reasonable $P({\rm TKER})$ and $P(N)$ given the considerably decreased
parametrization effort compared with TL-based PhysNet. On the other
hand, the roaming pathway can not be described realistically using the
present MS-ARMD model. The rates following overtone excitation of the
CH-stretch show the correct energy dependence compared with experiment
but are larger by one to two orders of magnitude which is attributed
to the higher barrier for H-transfer in the first step between {\it
  syn-}CH$_3$COOH and VHP. For the O--O scission energy the present
work suggests a best estimate of $D_e = 22$ kcal/mol which compares
with $\sim 25$ kcal/mol from electronic structure calculations at the
CASPT2 level of theory.\cite{lester:2016} The undulations superimposed
on $P({\rm TKER})$ are spaced by $\sim 1000$ cm$^{-1}$ which can be
potentially related to the C-C stretch mode in CH$_2$CHO which is
reported at 917 cm$^{-1}$.\cite{nistIR} Such an assignment is further
corroborated by separately analyzing the C-C distribution functions
for $\sim 100$ trajectories with the lowest- and highest-TKER from
Figure \ref{fig:tker}B, respectively. For large kinetic energy
release, less energy is available to distribute over the internal
degrees of freedom which leads to a smaller average C-C separation and
a more narrow distribution, whereas for low-TKER more energy is
available which shifts the average C-C separation to larger values and
widens up the distribution, see Figure S13.\\

\noindent
OH-roaming times until recombination to GA occurs are in the range of
picoseconds, consistent with earlier reports on roaming in
nitromethane.\cite{bowman:2014} Glycolaldehyde formation from
photodissociation of {\it syn-}CH$_3$COOH is therefore another case
for ``molecular roamers" (here OH).\cite{bowman:2014} As was reported
earlier, OH elimination at atmospheric conditions takes place on
longer ($\mu$s) time scales.\cite{fang:2016} Because OH-roaming to
form GA occurs on considerably shorter (ps) time scales, the processes
considered in the present work are also likely to be relevant to the
chemical evolution of the atmosphere, depending on the environmental
conditions (local density), and will lead to a rich and largely
unexplored chemistry. \\

\noindent
A notable finding of the present work is that the final distributions
for translational kinetic energy and rotational quantum numbers of the
OH fragment are sensitive to the O--O dissociation energy whereas the
energy-dependent rates $k(E)$ calculated using stretched exponential
function are not. Assuming $k(E)$ to follow an Arrhenius-type
expression at $T=300$ K would lead to an expected increase in the rate
by a factor of $\sim 150$, i.e. by approximately two orders of
magnitude, upon reducing the dissociation energy by 3 kcal/mol. This
is evidently not what is found from the present simulations which
rather lead to a change in the rate by a factor of 2 to 5, see Figure
\ref{fig:rate}. Therefore, elimination of OH from VHP is a {\it
  nonequilibrium process} because the reaction is initiated through
vibrational excitation of the CH-stretch and vibrational energy is
only equilibrated across all degrees of freedom on the time scale of
the reaction, and $k(E)$ is not particularly sensitive to the O--O
dissociation energy. This also suggests that the entire pathway from
the initial reactant (here {\it syn-}CH$_3$COOH) to the products
(here: CH$_2$CHO+OH, CH$_2$OHCO+H, and CH$_2$OH+HCO) needs to be
followed in order to realistically (re)distribute internal and
relative energies in the species involved.\\

\noindent
Based on the agreement for the final state distributions, $D_e \in
[22,25]$ kcal/mol is the recommended range from the present work. With
yet larger values for the O--O scission energy (e.g. $D_e^{\rm OO} =
28$ kcal/mol as in Figure S14) in particular $P({\rm
  TKER})$ deviate appreciably from the observations. Hence, a
combination of experimental results and advanced computational
modeling based on statistically significant numbers of trajectories
using a PES based on CASPT2 calculations allows to approximately
determine an essential quantity - the O--O scission energy - for
atmospheric modeling. This is comparable to estimating the
dissociation energy for HO$_3$ from energy-dependent rates and
statistical mechanical modeling,\cite{smith:2010} however, without
assuming thermal equilibrium in the present case. It is further noted
that the amount of roaming may be overestimated in the present work as
seen for $P({\rm TKER})$ in Figure \ref{fig:tker} which can be a
consequence of the rather small basis set used in the CASPT2
calculations and details in the long range interactions that can be
improved further.\\

\noindent
The current work also clarifies that a wide range of final products
are potentially accessible from a single initial reactant such as {\it
  syn-}CH$_3$COOH. Combining the methods used in the present work with
techniques to enumerate all possible fragmentation products of a
molecule,\cite{domin:2020} a comprehensive and quantitative
exploration of the decomposition dynamics of important atmospheric
molecules is in reach. The reactive radicals generated in such
processes engage in various bimolecular reactions, yielding outcomes
that have potentially significant consequences for atmospheric
modeling. The present study uncovers probably only a small portion of
the extensive possibilities in reactive intermediates and final
decomposition products of CIs accessible under atmospheric
conditions. It is likely that the findings of the present work are
also relevant for decomposition reactions for compounds other than
CIs. With recent advances to generate accurate, high-dimensional
reactive PESs\cite{unke:2021,MM.nnpes:2023} paired with
transfer-learning strategies\cite{MM.tl:2022,MM.tl:2023} and
integration into broadly applicable molecular simulation
software,\cite{MM.pycharmm:2023} routine exploration and discovery of
the decomposition dynamics of small to medium-sized molecules becomes
possible.\\

\section{Methods}

\subsection{Reactive Potential Energy Surfaces}
From previous work, reactive PESs are available as multi
state-adiabatic reactive molecular dynamics (MS-ARMD) and neural
network (PhysNet)
representations\cite{MM.criegee:2021,MM.armd:2014,MM.physnet:2019}
which both employed reference calculations at the MP2 level of theory.
The MS-ARMD PES fit to MP2/6-31G(d) reference data correctly describes
the barrier for H-transfer with a barrier height of 16.0 kcal/mol,
consistent with experimental estimations\cite{liu:2014} but the
dissociation energy for OH-elimination from VHP, for which no
experimental data is available, is considerably larger (31.5 kcal/mol)
compared with estimates between 18 and 26 kcal/mol from previous
multi-reference calculations.\cite{lester:2016,kurten:2012}
Consequently, the O--O scission energy in MS-ARMD was empirically
adjusted to match estimates of around 25 kcal/mol. Due to the
limitations of the MP2 reference calculations the PhysNet model also
suffers for the OH elimination step.\\

\noindent
Analysis of the two available PESs motivated the following specific
improvements. The PhysNet PES is transfer
learned\cite{smith:2019,MM.tl:2022,MM.tl:2023} to the
CASPT2(12,10)/cc-pVDZ level of theory based on reference energies,
forces and dipole moments for 26,500 structures computed by MOLPRO
2019.\cite{molpro:2020} The reference data set contains 16000
structures randomly chosen between CH$_3$COOH and OH elimination from
the MP2 data set and additional 10500 OH roaming, glycolaldehyde and
CH$_2$OH + CHO structures. The dissociation barrier for the O--O bond
at the CASPT2(12,10)/cc-pVDZ level of theory starting from the VHP
equilibrium to a separated vinoxy and OH radical complex is determined
with $\sim 22$ kcal/mol and is consistent with earlier
work.\cite{lester:2016} The quality of the transfer-learned PES is
shown in Figure \ref{sifig:pes}. For the MS-ARMD PES the O--O scission
energies considered were $D_e = 22, 25, 27$ kcal/mol and the charges
on the separating OH fragment were reduced to $q_{\rm O} = -0.1e ,
q_{\rm H} = 0.1e$ such that the PES correctly dissociates for large
O--O separations.\\

\noindent
In order to determine the sensitivity of the results from the MD
simulations on the O--O dissociation energy, the PhysNet model was
modified to increase the scission energy to $D_e \sim 25$ kcal/mol.
The modification adds an energy term $f({r_{\rm OO}})= \frac{\Delta
  V_s}{1 + e^{-b(r_{\rm OO}-r_e)}}$ to the transfer-learned PhysNet
representation to increase the dissociation energy by $\Delta
V_s$. The sigmoidal function depends on the O--O separation $r_{\rm
  OO}$ and the parameters $\Delta V_s = 3$ kcal/mol, $b = 14.0$
\AA\/$^{-1}$, $r_e = 1.95$ \AA\/ increase the dissociation energy and
influence the slope and the turning point of the curve,
respectively. For {\it syn-}CH$_3$COOH and VHP conformations
$f({r_{\rm OO}}) \sim 0$ which increases to $\Delta V_\mathrm{s} = 3$
kcal/mol as the OH radical dissociates.\\

\subsection{Molecular Dynamics Simulations}
Molecular dynamics (MD) simulations using the MS-ARMD PES were carried
out with a suitably extended c47a2 version of the CHARMM
program\cite{charmm:2009,MM.armd:2014} whereas simulations using the
PhysNet PES were carried out with the pyCHARMM API linked to c47a2 to
pipe the NN-energies and forces into the MD
code.\cite{pycharmm:2023,MM.pycharmm:2023}\\

\noindent
All production MD simulations, 200 ps in length for each run, were
carried out in the $NVE$ ensemble with a time step of $\Delta t = 0.1$
fs to conserve total energy. For each variant of the PESs (MS-ARMD and
PhysNet with corresponding values for $D_e^{\rm OO}$) at least 6000
independent trajectories were run. Initial conditions were generated
from $NVT$ simulations at 50 K from which structures and velocities
were extracted at regular intervals. Subsequently, the velocities of
the atoms involved in the C--H closest to the accepting oxygen O$_{\rm
  B}$ were excited by $\sim 2$ quanta along the bond vector equal to
excitation energies ranging from 5603 to 6082 cm$^{-1}$, consistent
with experiment.\cite{lester:2016} To emulate the cold conditions
encountered in the experiments the velocities of all remaining atoms
were set to zero.\\

\noindent
Simulations based on the MS-ARMD representation used this PES
throughout the entire simulation. With the PhysNet representation the
simulations used a hybrid protocol. For the H-transfer to form VHP
from {\it syn-}CH$_3$COOH the MS-ARMD representation featuring the
correct 16.0 kcal/mol barrier was used. Then, the positions $\vec{x}$
and momenta $\vec{p}$ of all atoms were stored and the simulation was
restarted using the PhysNet representation. For consistency, a scaling
$\lambda$ for the momenta $\vec{p}$ of all atoms was determined from
\begin{equation}
    T(\vec{p}) + \Delta V_\mathrm{MS-ARMD}(\vec{x}) = T(\lambda \cdot
    \vec{p}) + \Delta V_\mathrm{PhysNet}(\vec{x})
\end{equation}
to match the sum of the kinetic energy $T$ and the potential energy
difference $\Delta V$ between the potential at atom positions
$\vec{x}$ and the VHP equilibrium structure for both PESs.\\

\subsection{Final State Analysis and Rates}
For the final state analysis, the total energy of the separated system
was decomposed into fragment translational ($E_{\rm trans}$),
rotational ($E_{\rm rot}$), and vibrational ($E_{\rm vib}$) energy
components. The experimentally measured total kinetic energy release
(TKER) is the sum of the OH and vinoxy radical fragments translational
energy contributions $E_{{\rm trans},\alpha} = \vec{p}_{{\rm
    CM},\alpha}^{~2}/2M_\alpha$ where $M_\alpha$ is the mass and
$\vec{p}_{{\rm CM},\alpha}$ the center of mass momentum of fragment
$\alpha$ obtained as the sum of the respective atom momenta. The
rotational energy $E_{\rm rot}$ of the dissociating OH radical is
$E_{\rm rot} = L^2/2I$ with the angular momentum vector $\vec{L} =
\sum_{i=1}^{2} (\vec{r}_i - \vec{r}_{\rm CM}) \times \vec{p}_i$, where
$\vec{r}_i - \vec{r}_{\rm CM}$ is the atom position with respect to
the center of mass position, $\vec{p}_i$ are the atoms momenta and $I$
is the OH moment of inertia. As the exchange of energy between
vibrational and rotational degrees of freedom persists in the presence
of non-zero internal angular momentum, $E_{\rm rot}$ is averaged over
the final 25 ps after dissociation. The TKER and average $E_{\rm rot}$
become constant after both fragments have moved further than the
interaction cutoff of 6 \AA\/ from each other. The vibrational energy
of the OH fragment was then computed according to $E_{\rm vib} =
E_{\rm OH} - E_{\rm trans} - E_{\rm rot}$ with $E_{\rm OH} = T +
V(r_{\rm OH})$ which was invariably close to the quantum mechanical
ground state vibrational energy for the respective rotational state of
OH. Hence, final OH products are always in $\nu_{\rm OH} =
0$. Trajectories (30 \%) with OH-vibrational energies lower than the
quantum-mechanical zero-point energy at the respective rotational
state were excluded from the analysis.\\

\noindent
To determine reaction rates, for each energy $N_{\rm tot} = 6000$
individual simulations were carried out. The rates were determined
from fitting the number $N(t)$ of trajectories that had not reacted by
time $t$ to single ($\sim \exp(-kt)$) and stretched exponential decays
$(1-d) \exp(-kt)^{\gamma} + d)$, see Figure S9 for the
quality of the two fits.

\section*{Data Availability} 
The reference data and PhysNet codes that allow to reproduce the findings of this
study are openly available at
\url{https://github.com/MMunibas/Criegee-CASPT2 } and \url{https://github.com/MMunibas/PhysNet}.

\section*{Acknowledgment}
This work was partially supported by the Swiss National Science
Foundation through grants 200021-188724, the NCCR MUST (to MM), and
the University of Basel which is acknowledged. We also thank Prof. 
Michael N. R. Ashfold for valuable discussions.\\

\bibliography{refs.clean}

\providecommand{\latin}[1]{#1}
\makeatletter
\providecommand{\doi}
  {\begingroup\let\do\@makeother\dospecials
  \catcode`\{=1 \catcode`\}=2 \doi@aux}
\providecommand{\doi@aux}[1]{\endgroup\texttt{#1}}
\makeatother
\providecommand*\mcitethebibliography{\thebibliography}
\csname @ifundefined\endcsname{endmcitethebibliography}
  {\let\endmcitethebibliography\endthebibliography}{}
\begin{mcitethebibliography}{47}
\providecommand*\natexlab[1]{#1}
\providecommand*\mciteSetBstSublistMode[1]{}
\providecommand*\mciteSetBstMaxWidthForm[2]{}
\providecommand*\mciteBstWouldAddEndPuncttrue
  {\def\EndOfBibitem{\unskip.}}
\providecommand*\mciteBstWouldAddEndPunctfalse
  {\let\EndOfBibitem\relax}
\providecommand*\mciteSetBstMidEndSepPunct[3]{}
\providecommand*\mciteSetBstSublistLabelBeginEnd[3]{}
\providecommand*\EndOfBibitem{}
\mciteSetBstSublistMode{f}
\mciteSetBstMaxWidthForm{subitem}{(\alph{mcitesubitemcount})}
\mciteSetBstSublistLabelBeginEnd
  {\mcitemaxwidthsubitemform\space}
  {\relax}
  {\relax}

\bibitem[Vereecken \latin{et~al.}(2018)Vereecken, Aumont, Barnes, Bozzelli,
  Goldman, Green, Madronich, Mcgillen, Mellouki, Orlando, \latin{et~al.}
  others]{vereecken:2018}
Vereecken,~L.; Aumont,~B.; Barnes,~I.; Bozzelli,~J.; Goldman,~M.; Green,~W.;
  Madronich,~S.; Mcgillen,~M.; Mellouki,~A.; Orlando,~J. \latin{et~al.}
  Perspective on mechanism development and structure-activity relationships for
  gas-phase atmospheric chemistry. \emph{Int. J. Chem. Kin.} \textbf{2018},
  \emph{50}, 435--469\relax
\mciteBstWouldAddEndPuncttrue
\mciteSetBstMidEndSepPunct{\mcitedefaultmidpunct}
{\mcitedefaultendpunct}{\mcitedefaultseppunct}\relax
\EndOfBibitem
\bibitem[Rousso \latin{et~al.}(2019)Rousso, Hansen, Jasper, and
  Ju]{rousso:2019}
Rousso,~A.~C.; Hansen,~N.; Jasper,~A.~W.; Ju,~Y. Identification of the Criegee
  intermediate reaction network in ethylene ozonolysis: impact on energy
  conversion strategies and atmospheric chemistry. \emph{Phys. Chem. Chem.
  Phys.} \textbf{2019}, \emph{21}, 7341--7357\relax
\mciteBstWouldAddEndPuncttrue
\mciteSetBstMidEndSepPunct{\mcitedefaultmidpunct}
{\mcitedefaultendpunct}{\mcitedefaultseppunct}\relax
\EndOfBibitem
\bibitem[Criegee and Wenner(1949)Criegee, and Wenner]{criegee1949ozonisierung}
Criegee,~R.; Wenner,~G. Die Ozonisierung des 9, 10-Oktalins. \emph{Justus
  Liebigs Ann. Chem.} \textbf{1949}, \emph{564}, 9--15\relax
\mciteBstWouldAddEndPuncttrue
\mciteSetBstMidEndSepPunct{\mcitedefaultmidpunct}
{\mcitedefaultendpunct}{\mcitedefaultseppunct}\relax
\EndOfBibitem
\bibitem[Alam \latin{et~al.}(2011)Alam, Camredon, Rickard, Carr, Wyche,
  Hornsby, Monks, and Bloss]{alam2011total}
Alam,~M.~S.; Camredon,~M.; Rickard,~A.~R.; Carr,~T.; Wyche,~K.~P.;
  Hornsby,~K.~E.; Monks,~P.~S.; Bloss,~W.~J. Total radical yields from
  tropospheric ethene ozonolysis. \emph{Phys. Chem. Chem. Phys.} \textbf{2011},
  \emph{13}, 11002--11015\relax
\mciteBstWouldAddEndPuncttrue
\mciteSetBstMidEndSepPunct{\mcitedefaultmidpunct}
{\mcitedefaultendpunct}{\mcitedefaultseppunct}\relax
\EndOfBibitem
\bibitem[Novelli \latin{et~al.}(2014)Novelli, Vereecken, Lelieveld, and
  Harder]{novelli2014direct}
Novelli,~A.; Vereecken,~L.; Lelieveld,~J.; Harder,~H. Direct observation of OH
  formation from stabilised Criegee intermediates. \emph{Phys. Chem. Chem.
  Phys.} \textbf{2014}, \emph{16}, 19941--19951\relax
\mciteBstWouldAddEndPuncttrue
\mciteSetBstMidEndSepPunct{\mcitedefaultmidpunct}
{\mcitedefaultendpunct}{\mcitedefaultseppunct}\relax
\EndOfBibitem
\bibitem[Taatjes(2017)]{taatjes2017criegee}
Taatjes,~C.~A. Criegee intermediates: What direct production and detection can
  teach us about reactions of carbonyl oxides. \emph{Ann. Rev. Phys. Chem.}
  \textbf{2017}, \emph{68}, 183--207\relax
\mciteBstWouldAddEndPuncttrue
\mciteSetBstMidEndSepPunct{\mcitedefaultmidpunct}
{\mcitedefaultendpunct}{\mcitedefaultseppunct}\relax
\EndOfBibitem
\bibitem[Mauldin~Iii \latin{et~al.}(2012)Mauldin~Iii, Berndt, Sipil{\"a},
  Paasonen, Pet{\"a}j{\"a}, Kim, Kurt{\'e}n, Stratmann, Kerminen, and
  Kulmala]{mauldin2012new}
Mauldin~Iii,~R.; Berndt,~T.; Sipil{\"a},~M.; Paasonen,~P.; Pet{\"a}j{\"a},~T.;
  Kim,~S.; Kurt{\'e}n,~T.; Stratmann,~F.; Kerminen,~V.-M.; Kulmala,~M. A new
  atmospherically relevant oxidant of sulphur dioxide. \emph{Nature}
  \textbf{2012}, \emph{488}, 193--196\relax
\mciteBstWouldAddEndPuncttrue
\mciteSetBstMidEndSepPunct{\mcitedefaultmidpunct}
{\mcitedefaultendpunct}{\mcitedefaultseppunct}\relax
\EndOfBibitem
\bibitem[Welz \latin{et~al.}(2012)Welz, Savee, Osborn, Vasu, Percival,
  Shallcross, and Taatjes]{welz:2012}
Welz,~O.; Savee,~J.~D.; Osborn,~D.~L.; Vasu,~S.~S.; Percival,~C.~J.;
  Shallcross,~D.~E.; Taatjes,~C.~A. Direct kinetic measurements of Criegee
  intermediate (CH$_2$OO) formed by reaction of CH$_2$I with O$_2$.
  \emph{Science} \textbf{2012}, \emph{335}, 204--207\relax
\mciteBstWouldAddEndPuncttrue
\mciteSetBstMidEndSepPunct{\mcitedefaultmidpunct}
{\mcitedefaultendpunct}{\mcitedefaultseppunct}\relax
\EndOfBibitem
\bibitem[Kidwell \latin{et~al.}(2016)Kidwell, Li, Wang, Bowman, and
  Lester]{lester:2016}
Kidwell,~N.~M.; Li,~H.; Wang,~X.; Bowman,~J.~M.; Lester,~M.~I. Unimolecular
  dissociation dynamics of vibrationally activated CH$_3$CHOO Criegee
  intermediates to OH radical products. \emph{Nat. Chem.} \textbf{2016},
  \emph{8}, 509--514\relax
\mciteBstWouldAddEndPuncttrue
\mciteSetBstMidEndSepPunct{\mcitedefaultmidpunct}
{\mcitedefaultendpunct}{\mcitedefaultseppunct}\relax
\EndOfBibitem
\bibitem[Liu \latin{et~al.}(2014)Liu, Beames, Petit, McCoy, and
  Lester]{liu:2014}
Liu,~F.; Beames,~J.~M.; Petit,~A.~S.; McCoy,~A.~B.; Lester,~M.~I.
  Infrared-driven unimolecular reaction of CH$_3$CHOO Criegee intermediates to
  OH radical products. \emph{Science} \textbf{2014}, \emph{345},
  1596--1598\relax
\mciteBstWouldAddEndPuncttrue
\mciteSetBstMidEndSepPunct{\mcitedefaultmidpunct}
{\mcitedefaultendpunct}{\mcitedefaultseppunct}\relax
\EndOfBibitem
\bibitem[Fang \latin{et~al.}({2016})Fang, Liu, Barber, Klippenstein, McCoy, and
  Lester]{fang:2016}
Fang,~Y.; Liu,~F.; Barber,~V.~P.; Klippenstein,~S.~J.; McCoy,~A.~B.;
  Lester,~M.~I. {Communication: Real time observation of unimolecular decay of
  Criegee intermediates to OH radical products}. \emph{J. Chem. Phys.}
  \textbf{{2016}}, \emph{{144}}, 061102\relax
\mciteBstWouldAddEndPuncttrue
\mciteSetBstMidEndSepPunct{\mcitedefaultmidpunct}
{\mcitedefaultendpunct}{\mcitedefaultseppunct}\relax
\EndOfBibitem
\bibitem[Upadhyay and Meuwly(2021)Upadhyay, and Meuwly]{MM.criegee:2021}
Upadhyay,~M.; Meuwly,~M. Thermal and Vibrationally Activated Decomposition of
  the syn-CH$_3$CHOO Criegee Intermediate. \emph{ACS Earth Space Chem.}
  \textbf{2021}, \emph{5}, 3396--3406\relax
\mciteBstWouldAddEndPuncttrue
\mciteSetBstMidEndSepPunct{\mcitedefaultmidpunct}
{\mcitedefaultendpunct}{\mcitedefaultseppunct}\relax
\EndOfBibitem
\bibitem[Gligorovski \latin{et~al.}(2015)Gligorovski, Strekowski, Barbati, and
  Vione]{gligorovski2015environmental}
Gligorovski,~S.; Strekowski,~R.; Barbati,~S.; Vione,~D. Environmental
  implications of hydroxyl radicals (•OH). \emph{Chem. Rev.} \textbf{2015},
  \emph{115}, 13051--13092\relax
\mciteBstWouldAddEndPuncttrue
\mciteSetBstMidEndSepPunct{\mcitedefaultmidpunct}
{\mcitedefaultendpunct}{\mcitedefaultseppunct}\relax
\EndOfBibitem
\bibitem[Levy(1971)]{levy1971normal}
Levy,~H. Normal atmosphere: Large radical and formaldehyde concentrations
  predicted. \emph{Science} \textbf{1971}, \emph{173}, 141--143\relax
\mciteBstWouldAddEndPuncttrue
\mciteSetBstMidEndSepPunct{\mcitedefaultmidpunct}
{\mcitedefaultendpunct}{\mcitedefaultseppunct}\relax
\EndOfBibitem
\bibitem[Stone \latin{et~al.}(2012)Stone, Whalley, and Heard]{stone:2012}
Stone,~D.; Whalley,~L.~K.; Heard,~D.~E. Tropospheric OH and HO$_2$ radicals:
  field measurements and model comparisons. \emph{Chem. Soc. Rev.}
  \textbf{2012}, \emph{41}, 6348--6404\relax
\mciteBstWouldAddEndPuncttrue
\mciteSetBstMidEndSepPunct{\mcitedefaultmidpunct}
{\mcitedefaultendpunct}{\mcitedefaultseppunct}\relax
\EndOfBibitem
\bibitem[Emmerson and Carslaw(2009)Emmerson, and Carslaw]{emmerson2009night}
Emmerson,~K.; Carslaw,~N. Night-time radical chemistry during the TORCH
  campaign. \emph{Atmos. Environ.} \textbf{2009}, \emph{43}, 3220--3226\relax
\mciteBstWouldAddEndPuncttrue
\mciteSetBstMidEndSepPunct{\mcitedefaultmidpunct}
{\mcitedefaultendpunct}{\mcitedefaultseppunct}\relax
\EndOfBibitem
\bibitem[Khan \latin{et~al.}(2018)Khan, Percival, Caravan, Taatjes, and
  Shallcross]{khan2018criegee}
Khan,~M.; Percival,~C.; Caravan,~R.; Taatjes,~C.; Shallcross,~D. Criegee
  intermediates and their impacts on the troposphere. \emph{Environ. Sci.:
  Process. Impacts} \textbf{2018}, \emph{20}, 437--453\relax
\mciteBstWouldAddEndPuncttrue
\mciteSetBstMidEndSepPunct{\mcitedefaultmidpunct}
{\mcitedefaultendpunct}{\mcitedefaultseppunct}\relax
\EndOfBibitem
\bibitem[Conrad \latin{et~al.}(2021)Conrad, Hansen, Jasper, Thomason,
  Hidaldo-Rodrigues, Treshock, and Popolan-Vaida]{conrad2021identification}
Conrad,~A.~R.; Hansen,~N.; Jasper,~A.~W.; Thomason,~N.~K.;
  Hidaldo-Rodrigues,~L.; Treshock,~S.~P.; Popolan-Vaida,~D.~M. Identification
  of the acetaldehyde oxide Criegee intermediate reaction network in the
  ozone-assisted low-temperature oxidation of trans-2-butene. \emph{Phys. Chem.
  Chem. Phys.} \textbf{2021}, \emph{23}, 23554--23566\relax
\mciteBstWouldAddEndPuncttrue
\mciteSetBstMidEndSepPunct{\mcitedefaultmidpunct}
{\mcitedefaultendpunct}{\mcitedefaultseppunct}\relax
\EndOfBibitem
\bibitem[Lee \latin{et~al.}(2006)Lee, Goldstein, Kroll, Ng, Varutbangkul,
  Flagan, and Seinfeld]{lee2006gas}
Lee,~A.; Goldstein,~A.~H.; Kroll,~J.~H.; Ng,~N.~L.; Varutbangkul,~V.;
  Flagan,~R.~C.; Seinfeld,~J.~H. Gas-phase products and secondary aerosol
  yields from the photooxidation of 16 different terpenes. \emph{J. Geophys.
  Res. Atmos.} \textbf{2006}, \emph{111}, D17305\relax
\mciteBstWouldAddEndPuncttrue
\mciteSetBstMidEndSepPunct{\mcitedefaultmidpunct}
{\mcitedefaultendpunct}{\mcitedefaultseppunct}\relax
\EndOfBibitem
\bibitem[Niki \latin{et~al.}(1981)Niki, Maker, Savage, and
  Breitenbach]{niki1981ftir}
Niki,~H.; Maker,~P.; Savage,~C.; Breitenbach,~L. An FTIR study of mechanisms
  for the HO radical initiated oxidation of C$_2$H$_4$ in the presence of NO:
  detection of glycolaldehyde. \emph{Chem. Phys. Lett.} \textbf{1981},
  \emph{80}, 499--503\relax
\mciteBstWouldAddEndPuncttrue
\mciteSetBstMidEndSepPunct{\mcitedefaultmidpunct}
{\mcitedefaultendpunct}{\mcitedefaultseppunct}\relax
\EndOfBibitem
\bibitem[Bertschi \latin{et~al.}(2003)Bertschi, Yokelson, Ward, Babbitt,
  Susott, Goode, and Hao]{bertschi2003trace}
Bertschi,~I.; Yokelson,~R.~J.; Ward,~D.~E.; Babbitt,~R.~E.; Susott,~R.~A.;
  Goode,~J.~G.; Hao,~W.~M. Trace gas and particle emissions from fires in large
  diameter and below ground biomass fuels. \emph{J. Geophys. Res. Atmos.}
  \textbf{2003}, \emph{108}, 8472\relax
\mciteBstWouldAddEndPuncttrue
\mciteSetBstMidEndSepPunct{\mcitedefaultmidpunct}
{\mcitedefaultendpunct}{\mcitedefaultseppunct}\relax
\EndOfBibitem
\bibitem[Unke and Meuwly(2019)Unke, and Meuwly]{MM.physnet:2019}
Unke,~O.~T.; Meuwly,~M. PhysNet: A Neural Network for Predicting Energies,
  Forces, Dipole Moments, and Partial Charges. \emph{J. Chem. Theory Comput.}
  \textbf{2019}, \emph{15}, 3678--3693\relax
\mciteBstWouldAddEndPuncttrue
\mciteSetBstMidEndSepPunct{\mcitedefaultmidpunct}
{\mcitedefaultendpunct}{\mcitedefaultseppunct}\relax
\EndOfBibitem
\bibitem[Nagy \latin{et~al.}(2014)Nagy, Yosa~Reyes, and Meuwly]{MM.armd:2014}
Nagy,~T.; Yosa~Reyes,~J.; Meuwly,~M. Multisurface Adiabatic Reactive Molecular
  Dynamics. \emph{J. Chem. Theory Comput.} \textbf{2014}, \emph{10},
  1366--1375\relax
\mciteBstWouldAddEndPuncttrue
\mciteSetBstMidEndSepPunct{\mcitedefaultmidpunct}
{\mcitedefaultendpunct}{\mcitedefaultseppunct}\relax
\EndOfBibitem
\bibitem[Lester(2023)]{lester:2023}
Lester,~M. \emph{private communication} \textbf{2023}, \relax
\mciteBstWouldAddEndPunctfalse
\mciteSetBstMidEndSepPunct{\mcitedefaultmidpunct}
{}{\mcitedefaultseppunct}\relax
\EndOfBibitem
\bibitem[Olgu{\'\i}n-Arias \latin{et~al.}(2021)Olgu{\'\i}n-Arias, Davis, and
  Guti{\'e}rrez]{olguin:2021}
Olgu{\'\i}n-Arias,~V.; Davis,~S.; Guti{\'e}rrez,~G. A general statistical model
  for waiting times until collapse of a system. \emph{Physica A} \textbf{2021},
  \emph{561}, 125198\relax
\mciteBstWouldAddEndPuncttrue
\mciteSetBstMidEndSepPunct{\mcitedefaultmidpunct}
{\mcitedefaultendpunct}{\mcitedefaultseppunct}\relax
\EndOfBibitem
\bibitem[Yosa~Reyes and Meuwly(2011)Yosa~Reyes, and Meuwly]{yosa:2011}
Yosa~Reyes,~J.; Meuwly,~M. Vibrationally Induced Dissociation of Sulfuric Acid
  (H$_2$SO$_4$). \emph{J. Phys. Chem. A} \textbf{2011}, \emph{115},
  14350--14360\relax
\mciteBstWouldAddEndPuncttrue
\mciteSetBstMidEndSepPunct{\mcitedefaultmidpunct}
{\mcitedefaultendpunct}{\mcitedefaultseppunct}\relax
\EndOfBibitem
\bibitem[Yosa~Reyes \latin{et~al.}(2014)Yosa~Reyes, Nagy, and
  Meuwly]{reyes.pccp.2014.msarmd}
Yosa~Reyes,~J.; Nagy,~T.; Meuwly,~M. Competitive reaction pathways in
  vibrationally induced photodissociation of H$_2$SO$_4$. \emph{Phys. Chem.
  Chem. Phys.} \textbf{2014}, \emph{16}, 18533--18544\relax
\mciteBstWouldAddEndPuncttrue
\mciteSetBstMidEndSepPunct{\mcitedefaultmidpunct}
{\mcitedefaultendpunct}{\mcitedefaultseppunct}\relax
\EndOfBibitem
\bibitem[Austin \latin{et~al.}(1975)Austin, Beeson, Eisenstein, Frauenfelder,
  and Gunsalus]{austin:1975}
Austin,~R.; Beeson,~K.; Eisenstein,~L.; Frauenfelder,~H.; Gunsalus,~I. Dynamics
  of ligand binding to myoglobin. \emph{Biochemistry} \textbf{1975}, \emph{14},
  5355--5373\relax
\mciteBstWouldAddEndPuncttrue
\mciteSetBstMidEndSepPunct{\mcitedefaultmidpunct}
{\mcitedefaultendpunct}{\mcitedefaultseppunct}\relax
\EndOfBibitem
\bibitem[Soloviov \latin{et~al.}(2016)Soloviov, Das, and Meuwly]{MM.mbno:2016}
Soloviov,~M.; Das,~A.~K.; Meuwly,~M. Structural Interpretation of Metastable
  States in Myoglobin--NO. \emph{Angew. Chem. Int. Ed.} \textbf{2016},
  \emph{55}, 10126--10130\relax
\mciteBstWouldAddEndPuncttrue
\mciteSetBstMidEndSepPunct{\mcitedefaultmidpunct}
{\mcitedefaultendpunct}{\mcitedefaultseppunct}\relax
\EndOfBibitem
\bibitem[Foley \latin{et~al.}(2021)Foley, Xie, Guo, and
  Suits]{foley2021orbiting}
Foley,~C.~D.; Xie,~C.; Guo,~H.; Suits,~A.~G. Orbiting resonances in
  formaldehyde reveal coupling of roaming, radical, and molecular channels.
  \emph{Science} \textbf{2021}, \emph{374}, 1122--1127\relax
\mciteBstWouldAddEndPuncttrue
\mciteSetBstMidEndSepPunct{\mcitedefaultmidpunct}
{\mcitedefaultendpunct}{\mcitedefaultseppunct}\relax
\EndOfBibitem
\bibitem[Meuwly and Hutson(1999)Meuwly, and Hutson]{MM99:nehf}
Meuwly,~M.; Hutson,~J.~M. Morphing ab initio potentials: a systematic study of
  {Ne--HF}. \emph{J. Chem. Phys.} \textbf{1999}, \emph{110}, 8338--8347\relax
\mciteBstWouldAddEndPuncttrue
\mciteSetBstMidEndSepPunct{\mcitedefaultmidpunct}
{\mcitedefaultendpunct}{\mcitedefaultseppunct}\relax
\EndOfBibitem
\bibitem[Bowman and Gazdy(1991)Bowman, and Gazdy]{bowman:1991}
Bowman,~J.~M.; Gazdy,~B. A simple method to adjust potential energy surfaces:
  Application to HCO. \emph{J. Chem. Phys.} \textbf{1991}, \emph{94},
  816--817\relax
\mciteBstWouldAddEndPuncttrue
\mciteSetBstMidEndSepPunct{\mcitedefaultmidpunct}
{\mcitedefaultendpunct}{\mcitedefaultseppunct}\relax
\EndOfBibitem
\bibitem[Linstrom and Mallard(2001)Linstrom, and Mallard]{nistIR}
Linstrom,~P.; Mallard,~W. NIST Chemistry WebBook-SRD 69. \emph{National
  Institute of Standards and Technology Standard Reference Database}
  \textbf{2001}, \relax
\mciteBstWouldAddEndPunctfalse
\mciteSetBstMidEndSepPunct{\mcitedefaultmidpunct}
{}{\mcitedefaultseppunct}\relax
\EndOfBibitem
\bibitem[Bowman(2014)]{bowman:2014}
Bowman,~J.~M. Roaming. \emph{Mol. Phys.} \textbf{2014}, \emph{112},
  2516--2528\relax
\mciteBstWouldAddEndPuncttrue
\mciteSetBstMidEndSepPunct{\mcitedefaultmidpunct}
{\mcitedefaultendpunct}{\mcitedefaultseppunct}\relax
\EndOfBibitem
\bibitem[Le~Picard \latin{et~al.}(2010)Le~Picard, Tizniti, Canosa, Sims, and
  Smith]{smith:2010}
Le~Picard,~S.~D.; Tizniti,~M.; Canosa,~A.; Sims,~I.~R.; Smith,~I.~W. The
  thermodynamics of the elusive HO$_3$ radical. \emph{Science} \textbf{2010},
  \emph{328}, 1258--1262\relax
\mciteBstWouldAddEndPuncttrue
\mciteSetBstMidEndSepPunct{\mcitedefaultmidpunct}
{\mcitedefaultendpunct}{\mcitedefaultseppunct}\relax
\EndOfBibitem
\bibitem[D{\'e}sesquelles \latin{et~al.}(2020)D{\'e}sesquelles, Van-Oanh,
  Thomas, and Domin]{domin:2020}
D{\'e}sesquelles,~P.; Van-Oanh,~N.-T.; Thomas,~S.; Domin,~D. Statistical
  molecular fragmentation: which parameters influence the branching ratios?
  \emph{Phys. Chem. Chem. Phys.} \textbf{2020}, \emph{22}, 3160--3172\relax
\mciteBstWouldAddEndPuncttrue
\mciteSetBstMidEndSepPunct{\mcitedefaultmidpunct}
{\mcitedefaultendpunct}{\mcitedefaultseppunct}\relax
\EndOfBibitem
\bibitem[Unke \latin{et~al.}(2021)Unke, Chmiela, Sauceda, Gastegger, Poltavsky,
  Sch\"utt, Tkatchenko, and M\"uller]{unke:2021}
Unke,~O.~T.; Chmiela,~S.; Sauceda,~H.~E.; Gastegger,~M.; Poltavsky,~I.;
  Sch\"utt,~K.~T.; Tkatchenko,~A.; M\"uller,~K.-R. Machine learning force
  fields. \emph{Chem. Rev.} \textbf{2021}, \emph{121}, 10142--10186\relax
\mciteBstWouldAddEndPuncttrue
\mciteSetBstMidEndSepPunct{\mcitedefaultmidpunct}
{\mcitedefaultendpunct}{\mcitedefaultseppunct}\relax
\EndOfBibitem
\bibitem[K{\"a}ser \latin{et~al.}(2023)K{\"a}ser, Vazquez-Salazar, Meuwly, and
  T{\"o}pfer]{MM.nnpes:2023}
K{\"a}ser,~S.; Vazquez-Salazar,~L.~I.; Meuwly,~M.; T{\"o}pfer,~K. Neural
  network potentials for chemistry: concepts, applications and prospects.
  \emph{Digital Discovery} \textbf{2023}, \emph{2}, 28--58\relax
\mciteBstWouldAddEndPuncttrue
\mciteSetBstMidEndSepPunct{\mcitedefaultmidpunct}
{\mcitedefaultendpunct}{\mcitedefaultseppunct}\relax
\EndOfBibitem
\bibitem[K\"aser \latin{et~al.}(2022)K\"aser, Richardson, and
  Meuwly]{MM.tl:2022}
K\"aser,~S.; Richardson,~J.~O.; Meuwly,~M. Transfer Learning for Affordable and
  High-Quality Tunneling Splittings from Instanton Calculations. \emph{J. Chem.
  Theory Comput.} \textbf{2022}, \emph{18}, 6840--6850\relax
\mciteBstWouldAddEndPuncttrue
\mciteSetBstMidEndSepPunct{\mcitedefaultmidpunct}
{\mcitedefaultendpunct}{\mcitedefaultseppunct}\relax
\EndOfBibitem
\bibitem[K\"aser and Meuwly(2023)K\"aser, and Meuwly]{MM.tl:2023}
K\"aser,~S.; Meuwly,~M. Transfer-Learned Potential Energy Surfaces: Towards
  Microsecond-Scale Molecular Dynamics Simulations in the Gas Phase at CCSD(T)
  Quality. \emph{J. Chem. Phys.} \textbf{2023}, \relax
\mciteBstWouldAddEndPunctfalse
\mciteSetBstMidEndSepPunct{\mcitedefaultmidpunct}
{}{\mcitedefaultseppunct}\relax
\EndOfBibitem
\bibitem[Song \latin{et~al.}(2023)Song, K{\"a}ser, T{\"o}pfer, Vazquez-Salazar,
  and Meuwly]{MM.pycharmm:2023}
Song,~K.; K{\"a}ser,~S.; T{\"o}pfer,~K.; Vazquez-Salazar,~L.~I.; Meuwly,~M.
  PhysNet Meets CHARMM: A Framework for Routine Machine Learning/Molecular
  Mechanics Simulations. \emph{arXiv preprint arXiv:2304.12973} \textbf{2023},
  \relax
\mciteBstWouldAddEndPunctfalse
\mciteSetBstMidEndSepPunct{\mcitedefaultmidpunct}
{}{\mcitedefaultseppunct}\relax
\EndOfBibitem
\bibitem[Kurt\'en and Donahue(2012)Kurt\'en, and Donahue]{kurten:2012}
Kurt\'en,~T.; Donahue,~N.~M. MRCISD studies of the dissociation of
  vinylhydroperoxide, CH$_2$CHOOH: There is a saddle point. \emph{J. Phys.
  Chem. A} \textbf{2012}, \emph{116}, 6823--6830\relax
\mciteBstWouldAddEndPuncttrue
\mciteSetBstMidEndSepPunct{\mcitedefaultmidpunct}
{\mcitedefaultendpunct}{\mcitedefaultseppunct}\relax
\EndOfBibitem
\bibitem[Smith \latin{et~al.}(2019)Smith, Nebgen, Zubatyuk, Lubbers, Devereux,
  Barros, Tretiak, Isayev, and Roitberg]{smith:2019}
Smith,~J.~S.; Nebgen,~B.~T.; Zubatyuk,~R.; Lubbers,~N.; Devereux,~C.;
  Barros,~K.; Tretiak,~S.; Isayev,~O.; Roitberg,~A.~E. Approaching coupled
  cluster accuracy with a general-purpose neural network potential through
  transfer learning. \emph{Nat. Comm.} \textbf{2019}, \emph{10}, 2903\relax
\mciteBstWouldAddEndPuncttrue
\mciteSetBstMidEndSepPunct{\mcitedefaultmidpunct}
{\mcitedefaultendpunct}{\mcitedefaultseppunct}\relax
\EndOfBibitem
\bibitem[Werner \latin{et~al.}({2020})Werner, Knowles, Manby, Black, Doll,
  Hesselmann, Kats, Koehn, Korona, Kreplin, Ma, Miller, Mitrushchenkov,
  Peterson, Polyak, Rauhut, and Sibaev]{molpro:2020}
Werner,~H.-J.; Knowles,~P.~J.; Manby,~F.~R.; Black,~J.~A.; Doll,~K.;
  Hesselmann,~A.; Kats,~D.; Koehn,~A.; Korona,~T.; Kreplin,~D.~A.
  \latin{et~al.}  {The Molpro quantum chemistry package}. \emph{J. Chem. Phys.}
  \textbf{{2020}}, \emph{{152}}, {144107}\relax
\mciteBstWouldAddEndPuncttrue
\mciteSetBstMidEndSepPunct{\mcitedefaultmidpunct}
{\mcitedefaultendpunct}{\mcitedefaultseppunct}\relax
\EndOfBibitem
\bibitem[Brooks \latin{et~al.}(2009)Brooks, Brooks, Mackerell, Nilsson,
  Petrella, Roux, Won, Archontis, Bartels, Boresch, Caflisch, Caves, Cui,
  Dinner, Feig, Fischer, Gao, Hodoscek, Im, Kuczera, Lazaridis, Ma,
  Ovchinnikov, Paci, Pastor, Post, Pu, Schaefer, Tidor, Venable, Woodcock, Wu,
  Yang, York, and Karplus]{charmm:2009}
Brooks,~B.~R.; Brooks,~C.~L.,~III; Mackerell,~A.~D.,~Jr.; Nilsson,~L.;
  Petrella,~R.~J.; Roux,~B.; Won,~Y.; Archontis,~G.; Bartels,~C.; Boresch,~S.
  \latin{et~al.}  CHARMM: the biomolecular simulation program. \emph{J. Comput.
  Chem.} \textbf{2009}, \emph{30}, 1545--1614\relax
\mciteBstWouldAddEndPuncttrue
\mciteSetBstMidEndSepPunct{\mcitedefaultmidpunct}
{\mcitedefaultendpunct}{\mcitedefaultseppunct}\relax
\EndOfBibitem
\bibitem[Buckner \latin{et~al.}(2023)Buckner, Liu, Chakravorty, Wu, Cervantes,
  Lai, and Brooks~III]{pycharmm:2023}
Buckner,~J.; Liu,~X.; Chakravorty,~A.; Wu,~Y.; Cervantes,~L.~F.; Lai,~T.~T.;
  Brooks~III,~C.~L. pyCHARMM: Embedding CHARMM Functionality in a Python
  Framework. \emph{J. Chem. Theory Comput.} \textbf{2023}, in print\relax
\mciteBstWouldAddEndPuncttrue
\mciteSetBstMidEndSepPunct{\mcitedefaultmidpunct}
{\mcitedefaultendpunct}{\mcitedefaultseppunct}\relax
\EndOfBibitem
\end{mcitethebibliography}


\providecommand{\latin}[1]{#1}
\makeatletter
\providecommand{\doi}
  {\begingroup\let\do\@makeother\dospecials
  \catcode`\{=1 \catcode`\}=2 \doi@aux}
\providecommand{\doi@aux}[1]{\endgroup\texttt{#1}}
\makeatother
\providecommand*\mcitethebibliography{\thebibliography}
\csname @ifundefined\endcsname{endmcitethebibliography}
  {\let\endmcitethebibliography\endthebibliography}{}
\begin{mcitethebibliography}{5}
\providecommand*\natexlab[1]{#1}
\providecommand*\mciteSetBstSublistMode[1]{}
\providecommand*\mciteSetBstMaxWidthForm[2]{}
\providecommand*\mciteBstWouldAddEndPuncttrue
  {\def\EndOfBibitem{\unskip.}}
\providecommand*\mciteBstWouldAddEndPunctfalse
  {\let\EndOfBibitem\relax}
\providecommand*\mciteSetBstMidEndSepPunct[3]{}
\providecommand*\mciteSetBstSublistLabelBeginEnd[3]{}
\providecommand*\EndOfBibitem{}
\mciteSetBstSublistMode{f}
\mciteSetBstMaxWidthForm{subitem}{(\alph{mcitesubitemcount})}
\mciteSetBstSublistLabelBeginEnd
  {\mcitemaxwidthsubitemform\space}
  {\relax}
  {\relax}

\bibitem[Upadhyay and Meuwly(2021)Upadhyay, and Meuwly]{MM.criegee:2021}
Upadhyay,~M.; Meuwly,~M. Thermal and Vibrationally Activated Decomposition of
  the syn-CH$_3$CHOO Criegee Intermediate. \emph{ACS Earth Space Chem.}
  \textbf{2021}, \emph{5}, 3396--3406\relax
\mciteBstWouldAddEndPuncttrue
\mciteSetBstMidEndSepPunct{\mcitedefaultmidpunct}
{\mcitedefaultendpunct}{\mcitedefaultseppunct}\relax
\EndOfBibitem
\bibitem[Kidwell \latin{et~al.}(2016)Kidwell, Li, Wang, Bowman, and
  Lester]{lester:2016}
Kidwell,~N.~M.; Li,~H.; Wang,~X.; Bowman,~J.~M.; Lester,~M.~I. Unimolecular
  dissociation dynamics of vibrationally activated CH$_3$CHOO Criegee
  intermediates to OH radical products. \emph{Nat. Chem.} \textbf{2016},
  \emph{8}, 509--514\relax
\mciteBstWouldAddEndPuncttrue
\mciteSetBstMidEndSepPunct{\mcitedefaultmidpunct}
{\mcitedefaultendpunct}{\mcitedefaultseppunct}\relax
\EndOfBibitem
\bibitem[Liu \latin{et~al.}(2014)Liu, Beames, Petit, McCoy, and
  Lester]{liu:2014}
Liu,~F.; Beames,~J.~M.; Petit,~A.~S.; McCoy,~A.~B.; Lester,~M.~I.
  Infrared-driven unimolecular reaction of CH$_3$CHOO Criegee intermediates to
  OH radical products. \emph{Science} \textbf{2014}, \emph{345},
  1596--1598\relax
\mciteBstWouldAddEndPuncttrue
\mciteSetBstMidEndSepPunct{\mcitedefaultmidpunct}
{\mcitedefaultendpunct}{\mcitedefaultseppunct}\relax
\EndOfBibitem
\bibitem[Kurt\'en and Donahue(2012)Kurt\'en, and Donahue]{kurten:2012}
Kurt\'en,~T.; Donahue,~N.~M. MRCISD studies of the dissociation of
  vinylhydroperoxide, CH$_2$CHOOH: There is a saddle point. \emph{J. Phys.
  Chem. A} \textbf{2012}, \emph{116}, 6823--6830\relax
\mciteBstWouldAddEndPuncttrue
\mciteSetBstMidEndSepPunct{\mcitedefaultmidpunct}
{\mcitedefaultendpunct}{\mcitedefaultseppunct}\relax
\EndOfBibitem
\end{mcitethebibliography}

\end{document}


\today

\section{The Potential Energy Surface}
\begin{figure}[H]
\centering \includegraphics[scale=0.4]{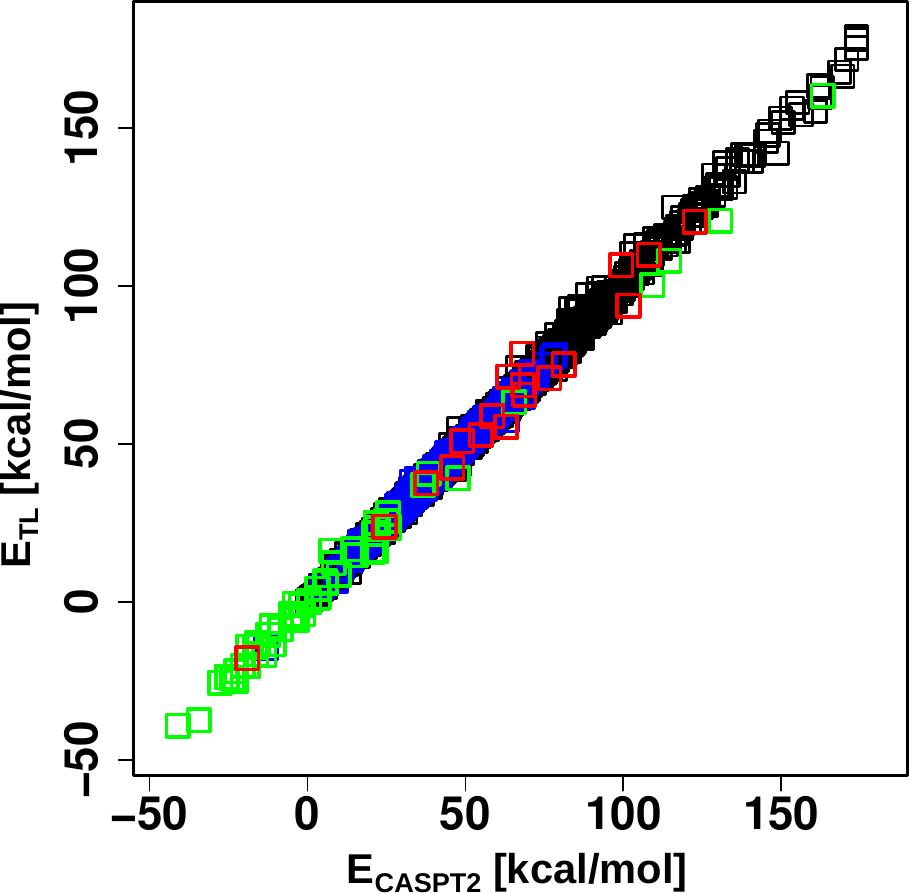}
    \caption{Correlation plot between the {\it ab initio} CASPT2
      energies and predicted TL energies for the 2750 test set
      structures from the entire reaction pathway.  The RMSE value for
      the predicted energies is 0.86 kcal/mol.  Color code:
      Glycolaldehyde (G; green), CH$_2$CHO + OH (D, E, F; blue), VHP
      and Criegee (A, B, C; black), HCO + CH$_2$OH (I; red).  Note
      that to CH$_2$OHCO+H was not part of the training.}
    \label{sifig:pes}
\end{figure}

The quality of the transfer-learned PES to the CASPT2(12,10)/cc-pVDZ
level of theory is reported in Figure \ref{sifig:pes}. Out of total
26500 structures, 21000 structures were used for training and 2750
each for validation and testing the TL model.\\

\noindent
First, the quality and characteristics of the PESs employed in the
present work are discussed and compared with what is known from the
literature. Experimentally, the barrier for H-transfer and the
dissociation energy of the O--O bond for OH fragmentation into
CH$_2$CHO+OH are important features. The transition state for
H-transfer from {\it syn-}CH$_3$CHOO to VHP was found at
16.0/18.6/19.8 kcal/mol at the
MP2\cite{MM.criegee:2021}/CCSD(T)\cite{lester:2016}/CASPT2 levels of
theory, respectively, compared with an experimentally reported barrier
of $\le$ 16.0 kcal/mol (5600 cm$^{-1}$) based on the lowest energy
leading to OH products\cite{liu:2014}. Because zero-point vibrational
energy is not present in the classical MD simulations and {\it
  syn-}Criegee is internally cold at the moment of CH-excitation, the
barrier of 16.0 kcal/mol from the MP2 calculations is consistent with
an experimental excitation energy separating the ground vibrational
state and the TS leading to VHP. For the O--O scission energy the
computed values are 31.5/18.8/26.2 kcal/mol at the
MP2/MRCISD(4,4)/CASPT2 levels of
theory.\cite{MM.criegee:2021,kurten:2012,lester:2016} However, no
information on this is available from experiments.\\

\noindent
The energy profile of all species considered in the present work at
the respective levels of theory (MP2/6-31G(d) for the {\it
  syn-}Criegee, VHP and the TS between them; CASPT2(12,10)/cc-pVDZ for
all remaining states) is reported in Figure 1. The
available parametrization based on MP2/aug-cc-pVTZ was not used
because the barrier for H-transfer (14.9 kcal/mol) underestimates the
value known from experiment. Transfer learning the original MP2
PhysNet model to the CASPT2(12,10)/cc-pVDZ level yields the
correlation plot shown in Figure \ref{sifig:pes}. Over a range of more
than 200 kcal/mol the root mean squared error for 2750 test structures
(not used in TL) is 0.86 kcal/mol for all species involved, which
establishes the high quality of the transfer-learned PES. Evaluating
the stationary points on the TL-PhysNet PES yields differences to
energies from electronic structure calculations of $\leq 0.3$ kcal/mol
except for the fragmentation products for GA, namely CH$_2$OHCO+H and
CH$_2$OH+HCO for which a $\sim 100$ structures were included in the
data set but the CH$_2$OHCO+H decay channel was not covered because
this reactive step was not of primary interest in the present work.\\

\section{Motivating the Value for $t_e$}
\begin{figure}
\centering \includegraphics[scale=0.3]{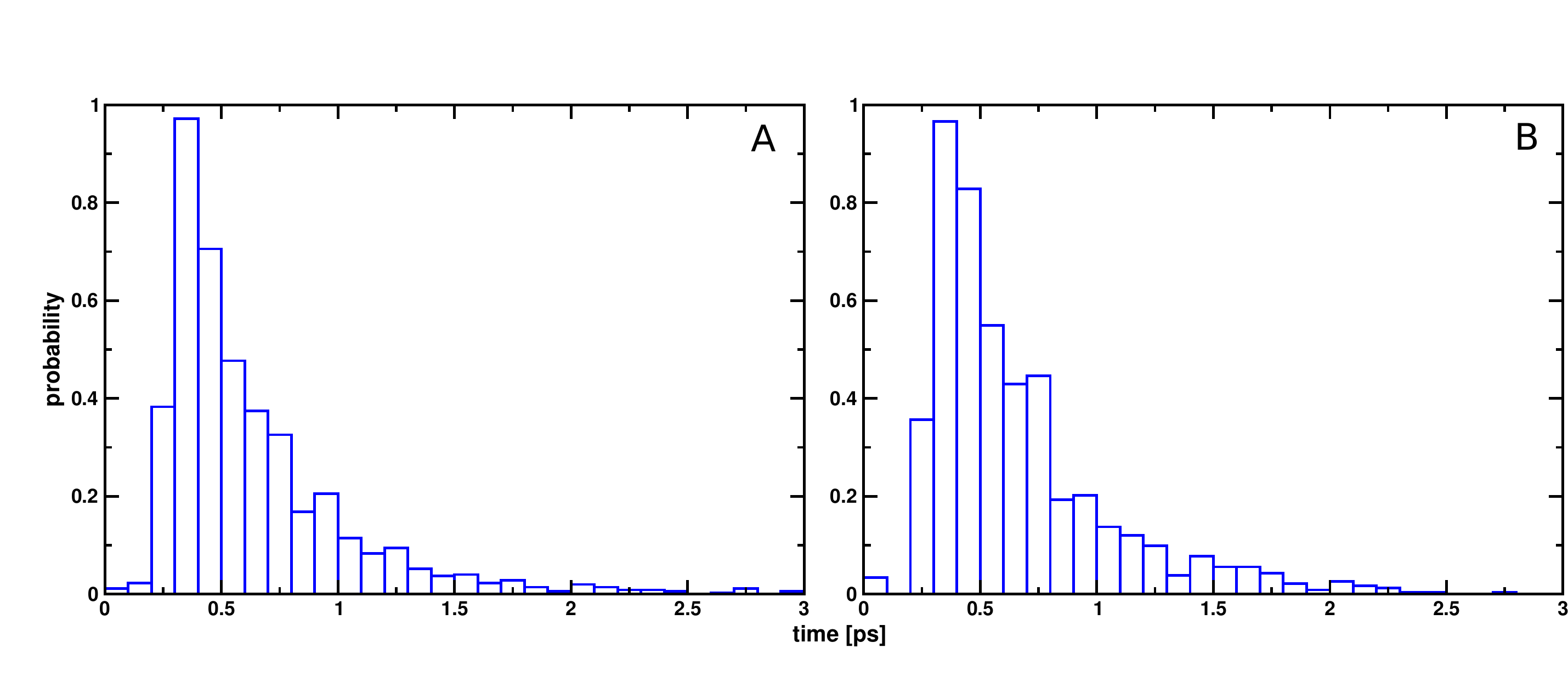}
   \caption{Distribution of OH dissociation times from O--O separation
     3 \AA\/ to 10 \AA\/ (for trajectories in which CH$_2$CHO and OH
     fragments are the reaction products). The time at which the O--O
     separation reaches 3 \AA\/ is set to be the zero of time. Panel A
     and B are for $D_e = 22$ and $D_e = 25$ kcal/mol respectively
     with CH-excitation at 5988 cm$^{-1}$ using the PhysNet PES.}
    \label{sifig:oh.time}
\end{figure}

\begin{figure}[H]
\centering \includegraphics[scale=0.45]{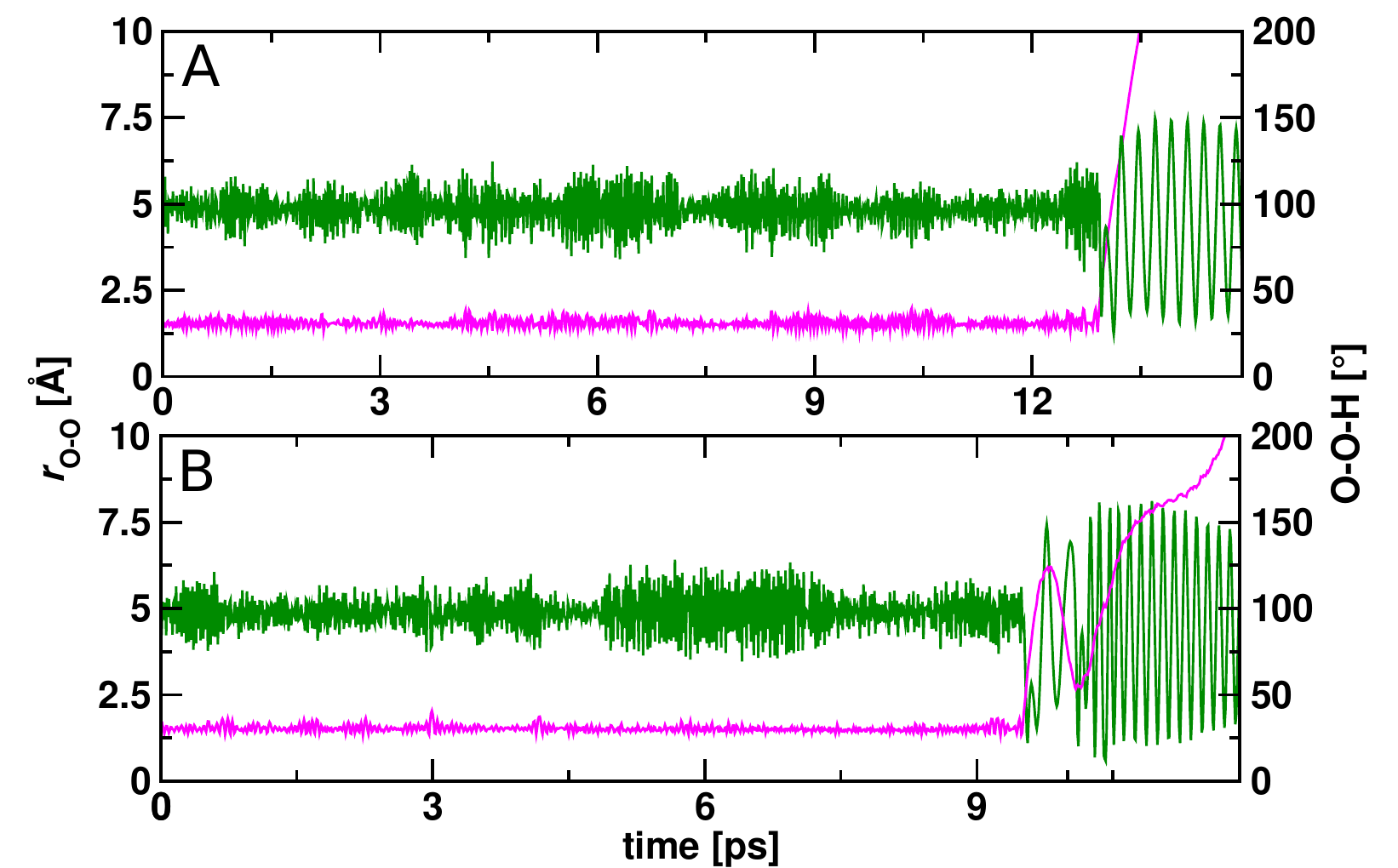}
   \caption{Time series for the O--O separation and the O--(OH) angle
     from a trajectory showing direct (Panel A) and roaming OH
     dissociation (Panel B) using PhysNet PES with $D_e = 22$ kcal/mol
     and CH-excitation at 5988 cm$^{-1}$.}
    \label{sifig:roo}
\end{figure}

\begin{figure}[H]
\centering \includegraphics[scale=0.50]{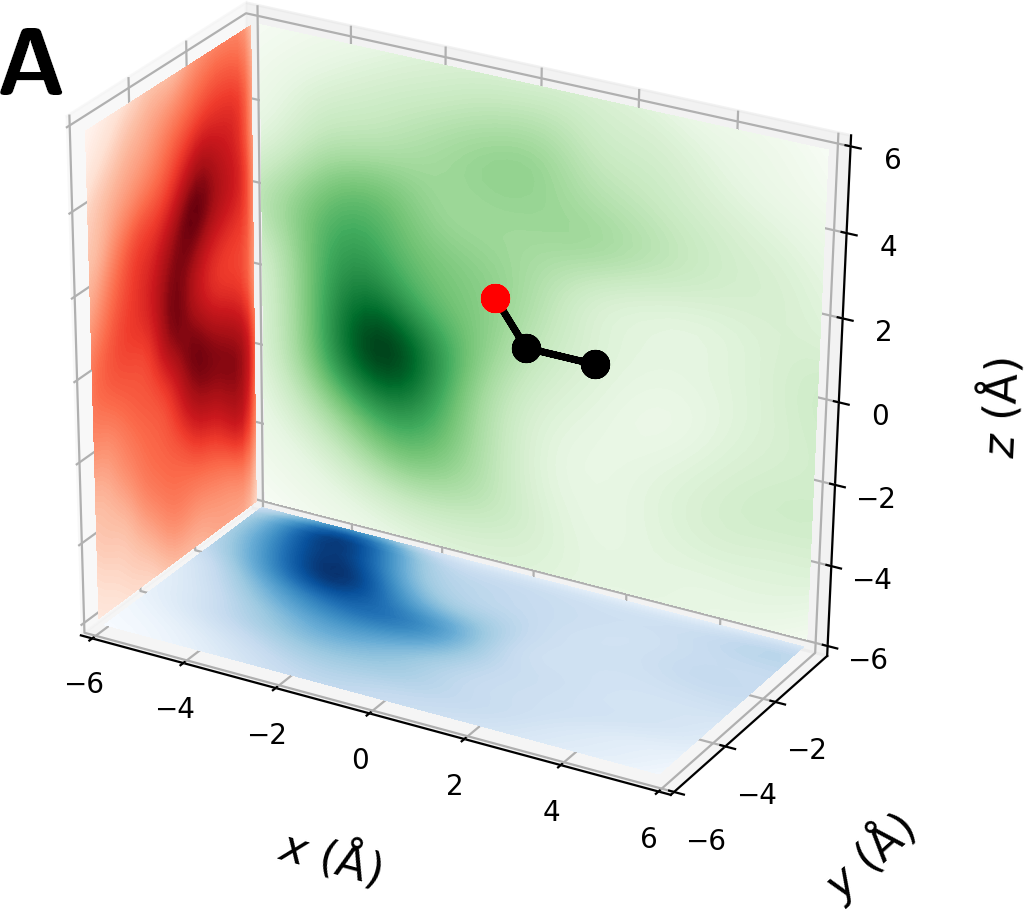}
\centering \includegraphics[scale=0.50]{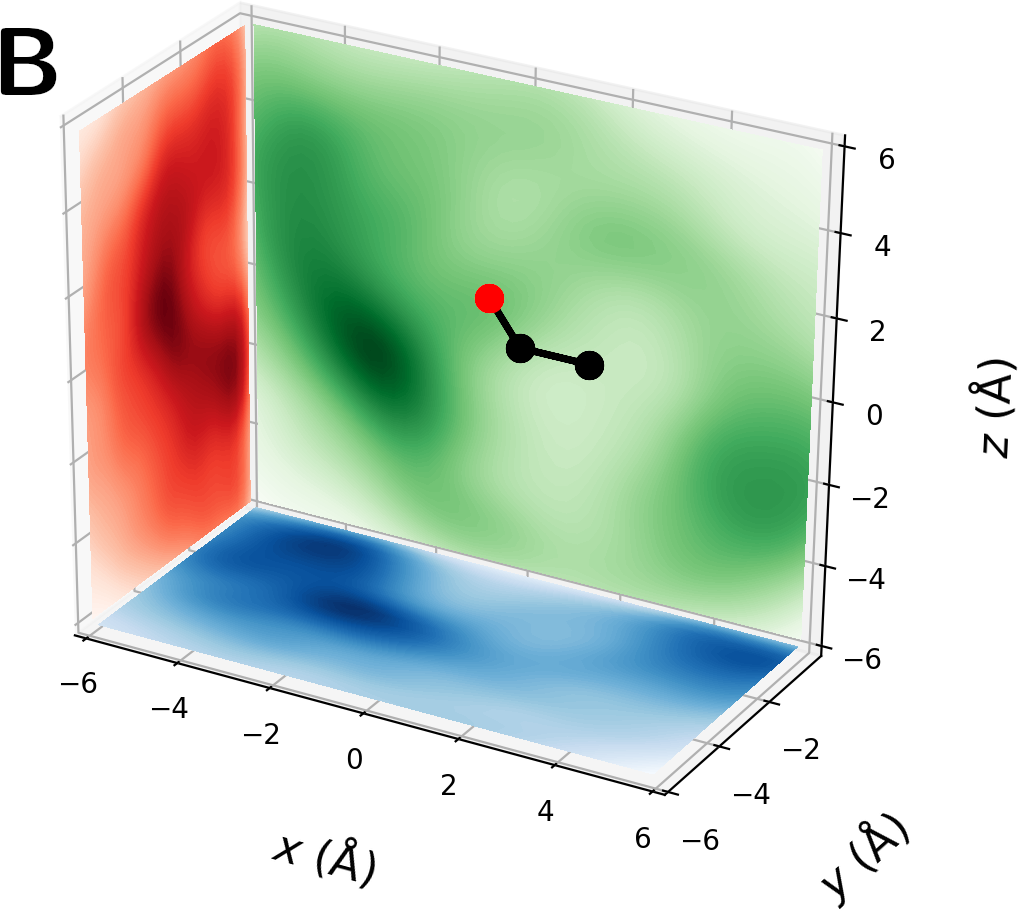}
   \caption{Probability densities of the OH radical around the
     CH$_{2}$CHO group (with C$_{\rm A}$-C$_{\rm B}$-O$_{\rm A}$ in
     the $xz$ plane and the center of mass of CH$_{2}$CHO in the
     origin) from $\sim 500$ trajectories each leading to A: direct or
     B: roaming OH dissociation on the PhysNet PES with $D_e = 22$
     kcal/mol. Trajectories in panel B access the half planes $(x,y)$
     and $(x,z)$ which are not visited in panel A for directly
     dissociating trajectories.}
    \label{sifig:density}
\end{figure}

To motivate the value of $t_e = 0.8$ ps, the distribution of
dissociation times using PhysNet PESs for $D_e= 22$ and 25 kcal/mol
are reported in Figure \ref{sifig:oh.time}. For both distributions, a
step at $t_e > 0.8$ ps is found. Furthermore, a monotonic increase in
the O--O separation with time is observed in the trajectories with
$t_e < 0.8$ ps, see Figure \ref{sifig:roo}A. On the other hand, for a
``roaming elimination'' shown in Figure \ref{sifig:roo}B the OH
fragment first reaches an O--O separation of $\sim 6$ \AA\/ and
re-approaches to 2.5 \AA\/ before finally dissociating from
CH$_2$CHO. For a positionally resolved picture of the different types
of OH dissociation, density maps of OH radical around CH$_2$CHO are
shown in Figure \ref{sifig:density}. Roaming trajectories (Figure
\ref{sifig:density}B) differ from direct dissociation (Figure
\ref{sifig:density}A) by the coverage of the surrounding space by the
OH radical around the vinoxy group. Unlike direct dissociation,
roaming OH explores the vicinity of the vinoxy radical before complete
fragmentation, see Figure \ref{sifig:density}.\\

\begin{figure}[H]
\centering \includegraphics[scale=0.3]{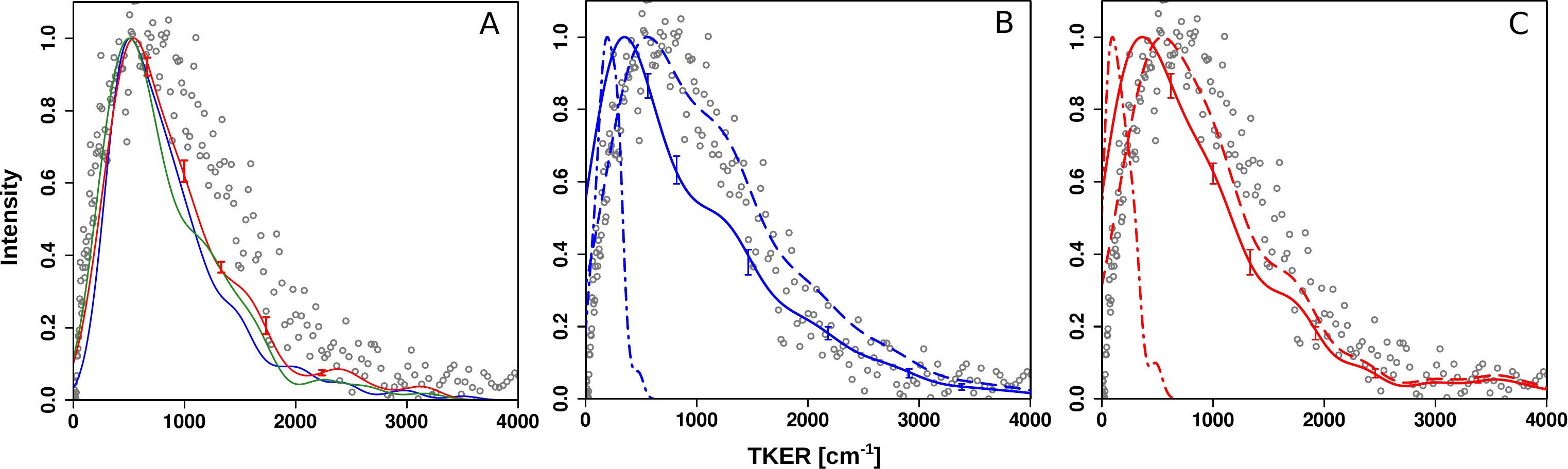}
   \caption{TKER forming OH(X$^2 \Pi$, $v=0$, $N=3$) from simulations
     (at 5988 cm$^{-1}$) and from experimental (grey
     circles)\cite{lester:2016} $P({\rm TKER})$ (at 6081 cm$^{-1}$)
     forming OH(X$^2 \Pi$, $v=0$, $N=3$). Panel A: MS-ARMD, Panel B:
     PhysNet with conventional $D_e = 22$ kcal/mol, Panel C: PhysNet
     with $D_e = 25$ kcal/mol. Solid, dotted-dashed and dashed lines
     correspond to all (direct+roaming), roaming and direct
     dissociated trajectories. Results are shown for $D_e = 22$
     kcal/mol (blue), $D_e = 25$ kcal/mol (red), and $D_e = 27$
     kcal/mol (green). For MS-ARMD all OH formed through direct
     dissociation. For formation of OH(X$^2 \Pi$, $v=0$, all $N$) see
     Figure 2 in the main MS.}
    \label{sifig:oh-n3}
\end{figure}

\begin{figure}[H]
    \centering \includegraphics[scale=0.52]{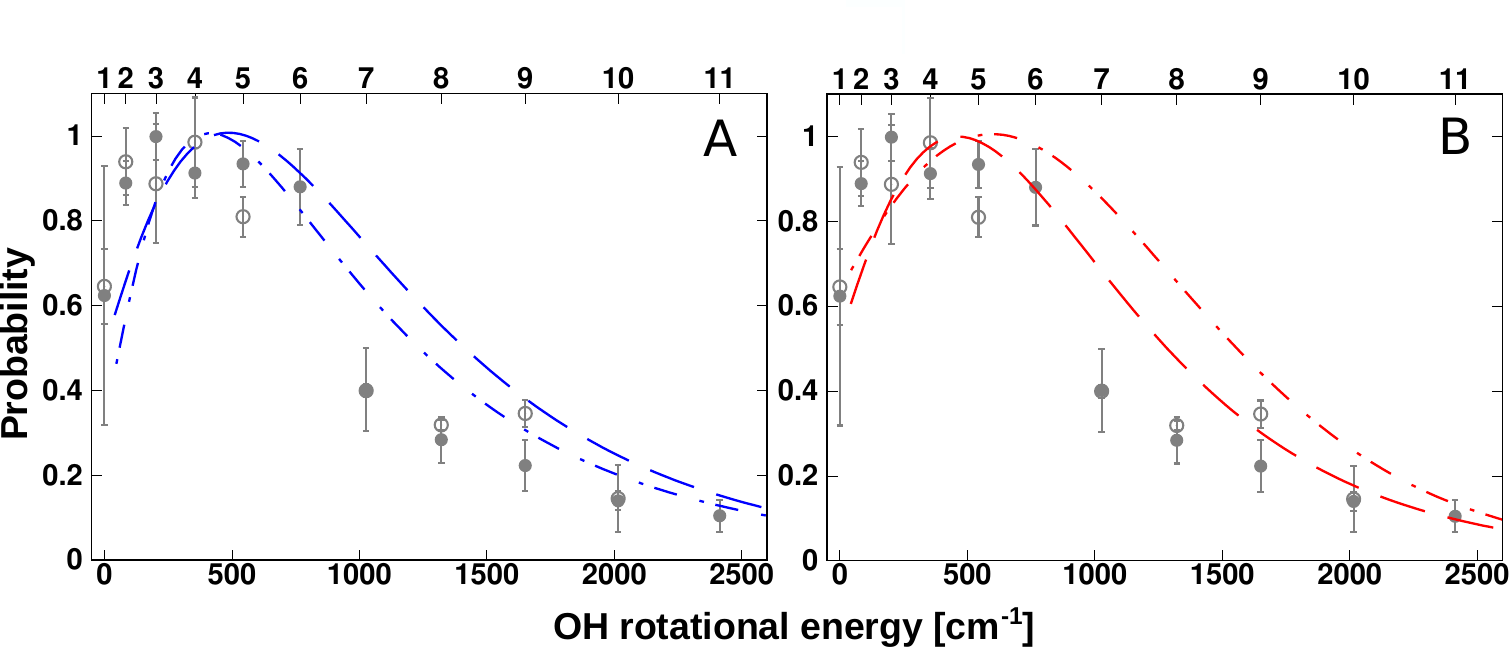}
    \caption{Rotational energy and corresponding state distributions
      with CH-excitation at 5988 cm$^{-1}$ using PhysNet PES with $D_e
      = 22$ kcal/mol (panel A) and $D_e = 25$ kcal/mol (panel B)
      compared with experiments (symbols). Here dotted-dashed and
      dashed lines correspond to roaming and directly dissociated
      trajectories. The most probable state is set to unity.}
    \label{sifig:4}
\end{figure}

\begin{figure}
    \centering \includegraphics[scale=0.42]{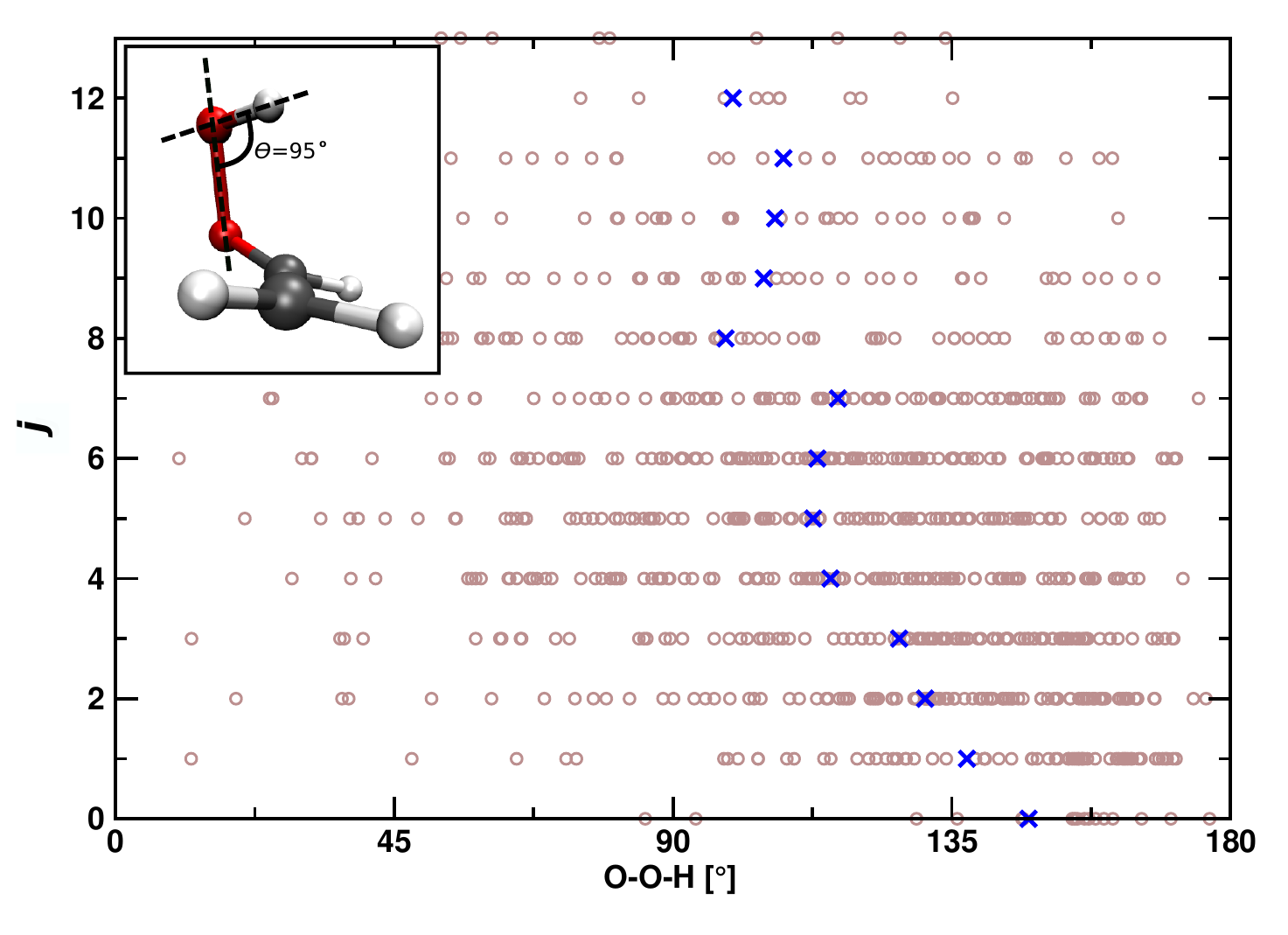}
    \caption{The final OH rotational quantum number $N$ vs.  the O-O-H
      angle at an O--O separation of 3 \AA\/ from 1170 independent
      simulations using the PhysNet PES with $D_e = 22$ kcal/mol and
      CH-excitation energy of 5988 cm$^{-1}$. The individual values
      are the brown circles and their average are the blue solid
      cross. The inset shows the definition of the OOH angle. The
      final $N-$state is quite well correlated with the average angle
      at the moment of dissociation.}
    \label{sifig:jvstheta}
\end{figure}

\begin{figure}
\centering \includegraphics[scale=0.4]{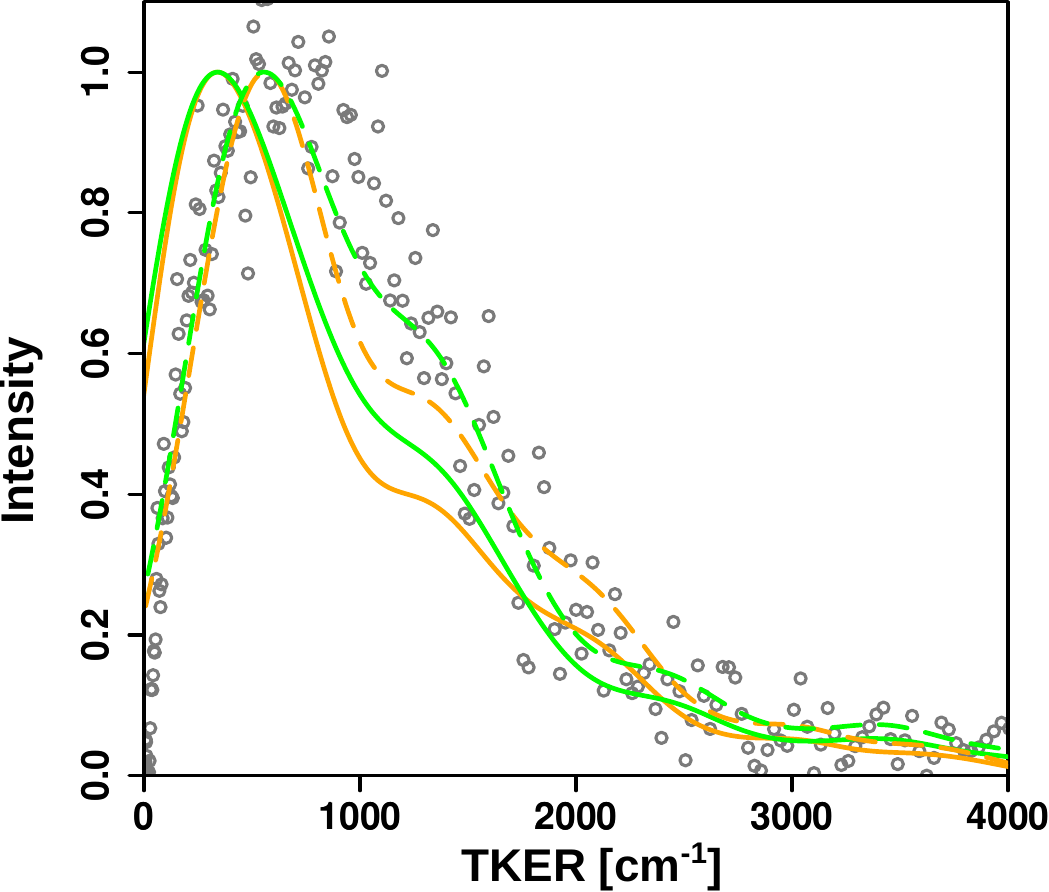}
   \caption{TKER forming OH(X$^2 \Pi$, $v=0$, $N \in [0,3]$) (orange)
     and OH(X$^2 \Pi$, $v=0$, $N \in [9-12]$) (green) from simulations
     (at 5988 cm$^{-1}$) and from experiments (grey
     circles)\cite{lester:2016} $P({\rm TKER})$ (at 6081 cm$^{-1}$)
     forming OH(X$^2 \Pi$, $v=0$, $N=3$) for $D_e = 22$
     kcal/mol. Solid lines for ``all trajectories'' and dashed for
     ``direct dissociation''.}
    \label{sifig:oh-lown-highn}
\end{figure}

\begin{figure}
\centering
    \includegraphics[scale=0.38]{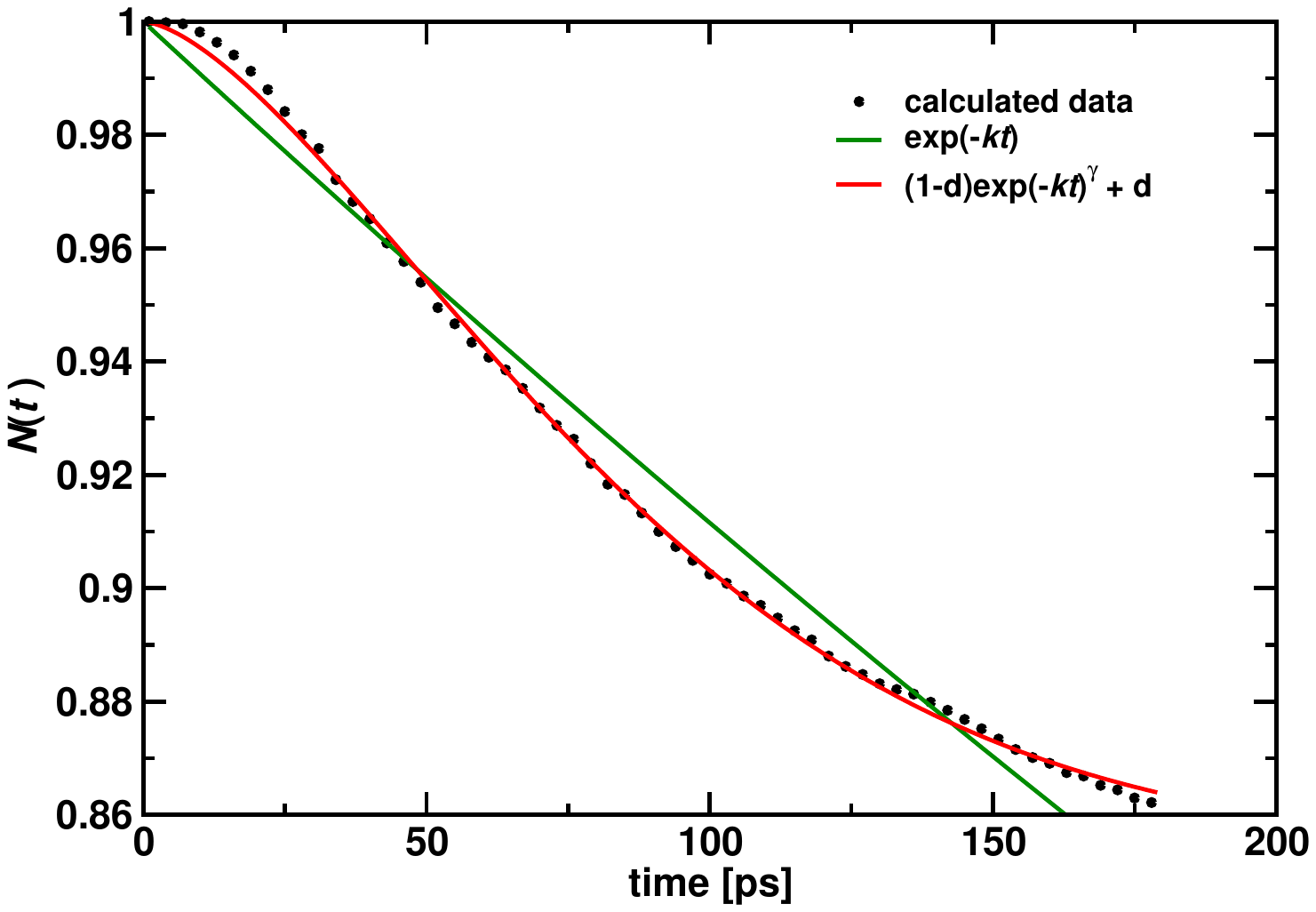}
   \caption{Fit of $N(t)$ (black circles) to single (green) and
     stretched (red) exponential for simulations using the PhysNet PES
     with $D_e= 22$ kcal/mol and vibrational excitation at 5603
     cm$^{-1}$.}
    \label{sifig:rate}
\end{figure}

\begin{figure}
\centering
    \includegraphics[scale=0.38]{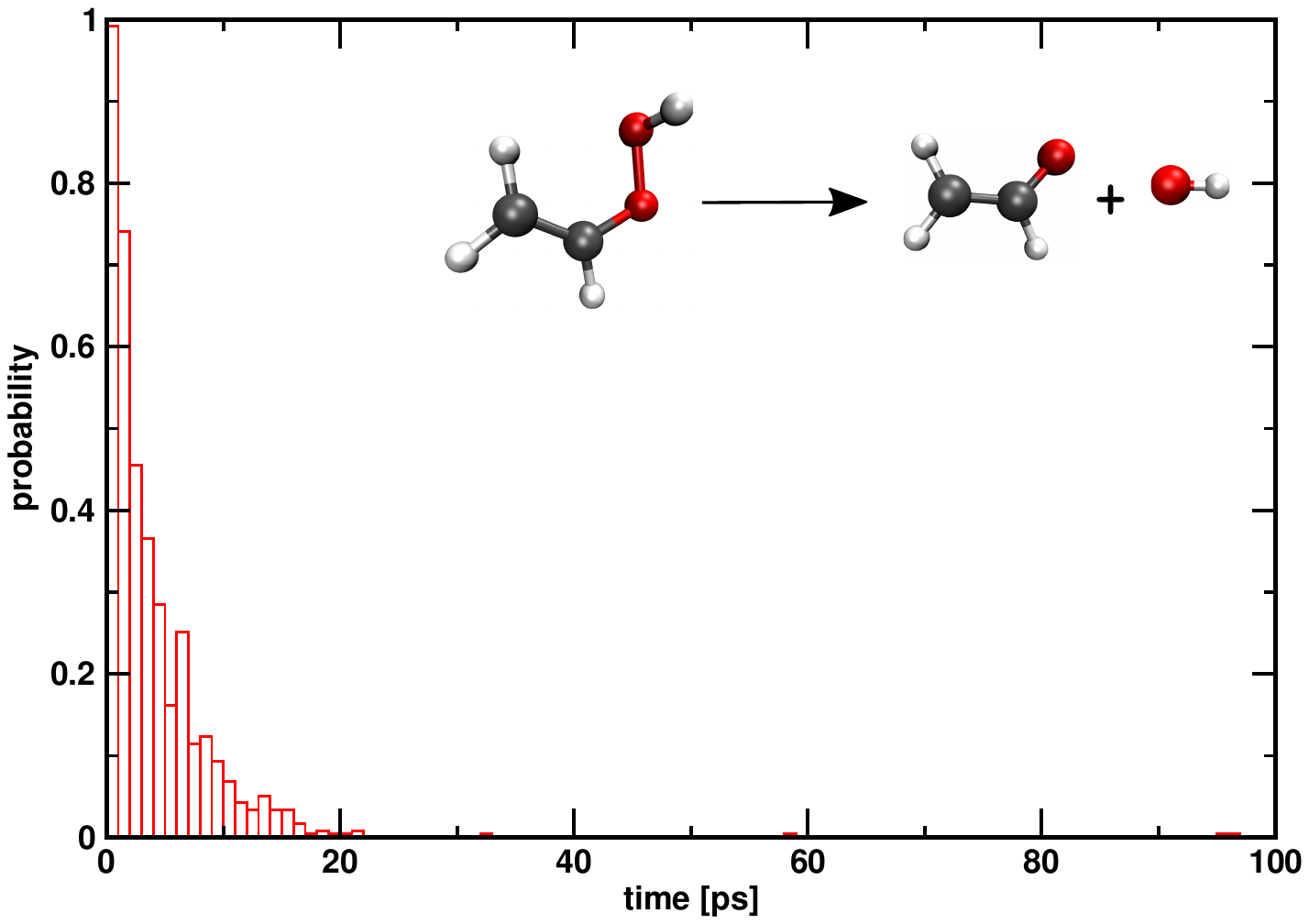}
   \caption{Lifetime distribution for VHP \meenu{before dissociating to CH$_2$CHO + OH} from simulations using the
     PhysNet PES with $D_e = 22$ kcal/mol and CH excitation at 5988
     cm$^{-1}$.}
    \label{sifig:vhplife}
\end{figure}

\begin{figure}
\centering
    \includegraphics[scale=0.38]{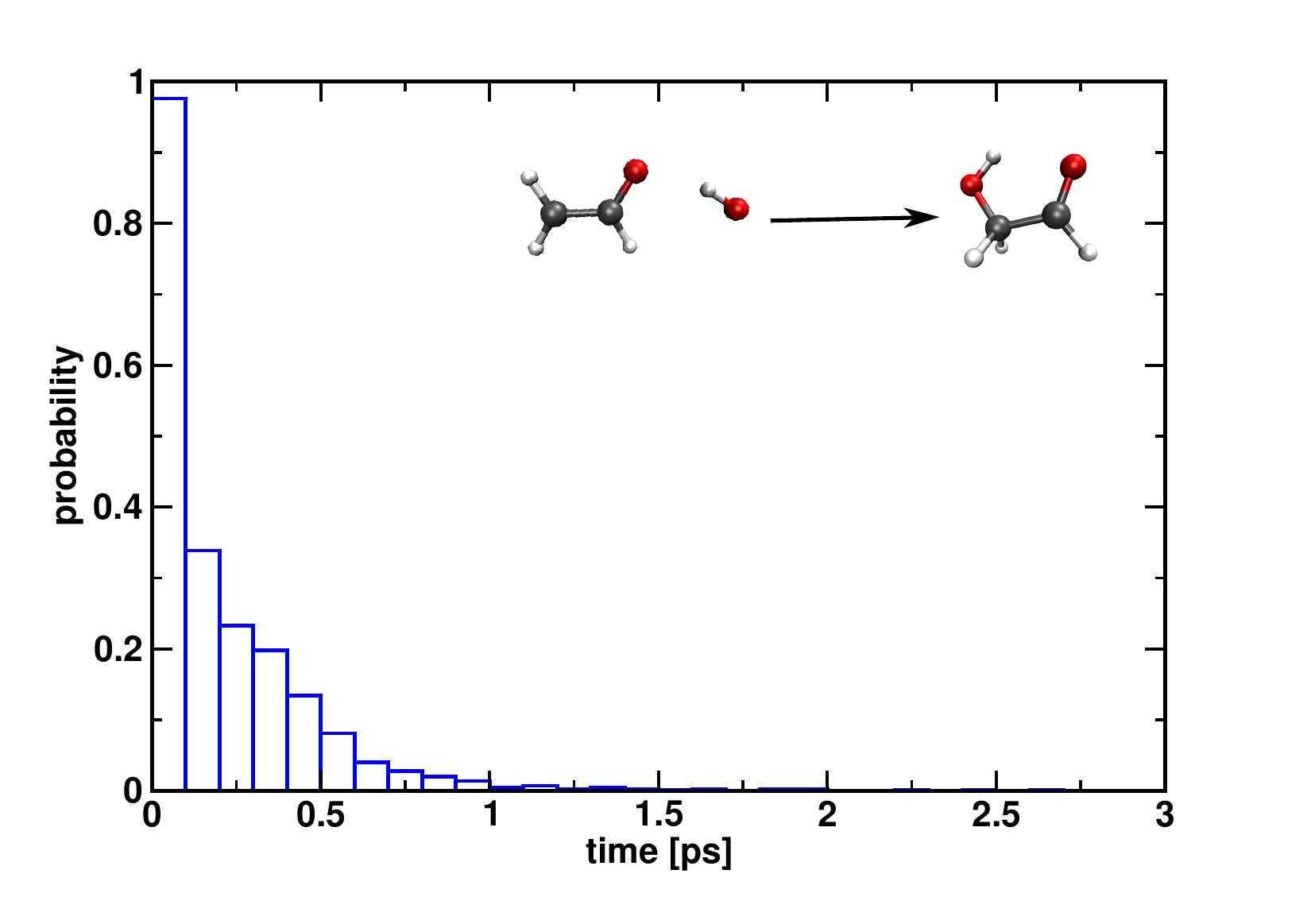}
   \caption{Distribution of reaction times for CH$_2$CHO + OH
     $\longrightarrow$ glycolaldehyde using the PhysNet PES with $D_e
     = 22$ kcal/mol and CH excitation at 5988 cm$^{-1}$. Here the time
     at which the O--O separation reaches 3 \AA\/ is set as zero. The
     pronounced peak at short reaction time ($\tau_r \leq 0.1$ ps)
     involves direct OH transfer to the CH$_2$ group whereas the
     remainder of the distribution is for OH roaming trajectories, see
     Figure 7A for a trajectory with $\tau_r = 1$ ps.}
    \label{sifig:ga.reaction}
\end{figure}

\begin{figure}
\centering
    \includegraphics[scale=0.38]{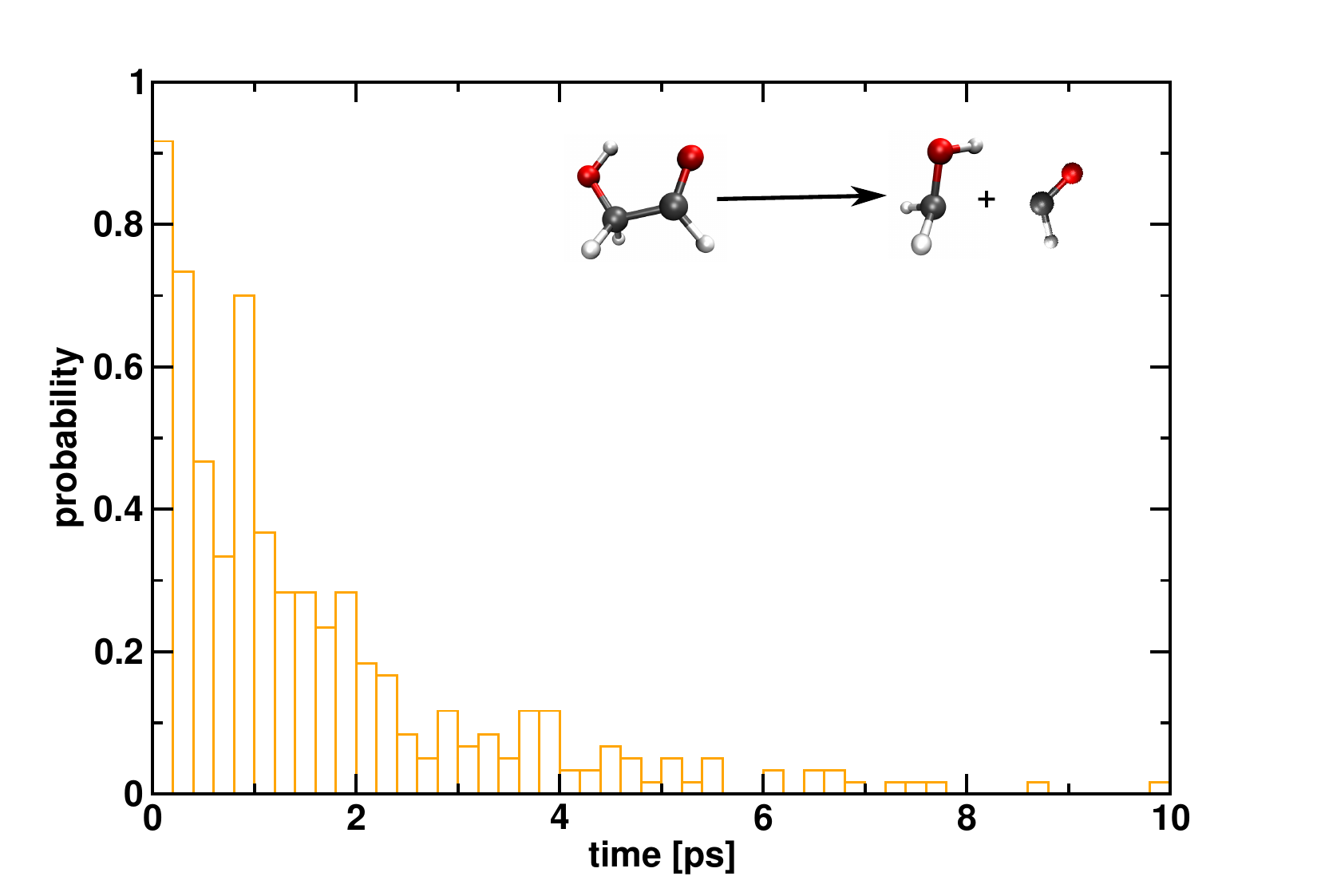}
   \caption{Lifetime distribution of glycolaldehyde before
     dissociating to CH$_2$OH + HCO using the PhysNet PES with $D_e =
     22$ kcal/mol and CH excitation at 5988 cm$^{-1}$.}
    \label{sifig:ga.life}
\end{figure}

\begin{figure}
\centering \includegraphics[scale=0.38]{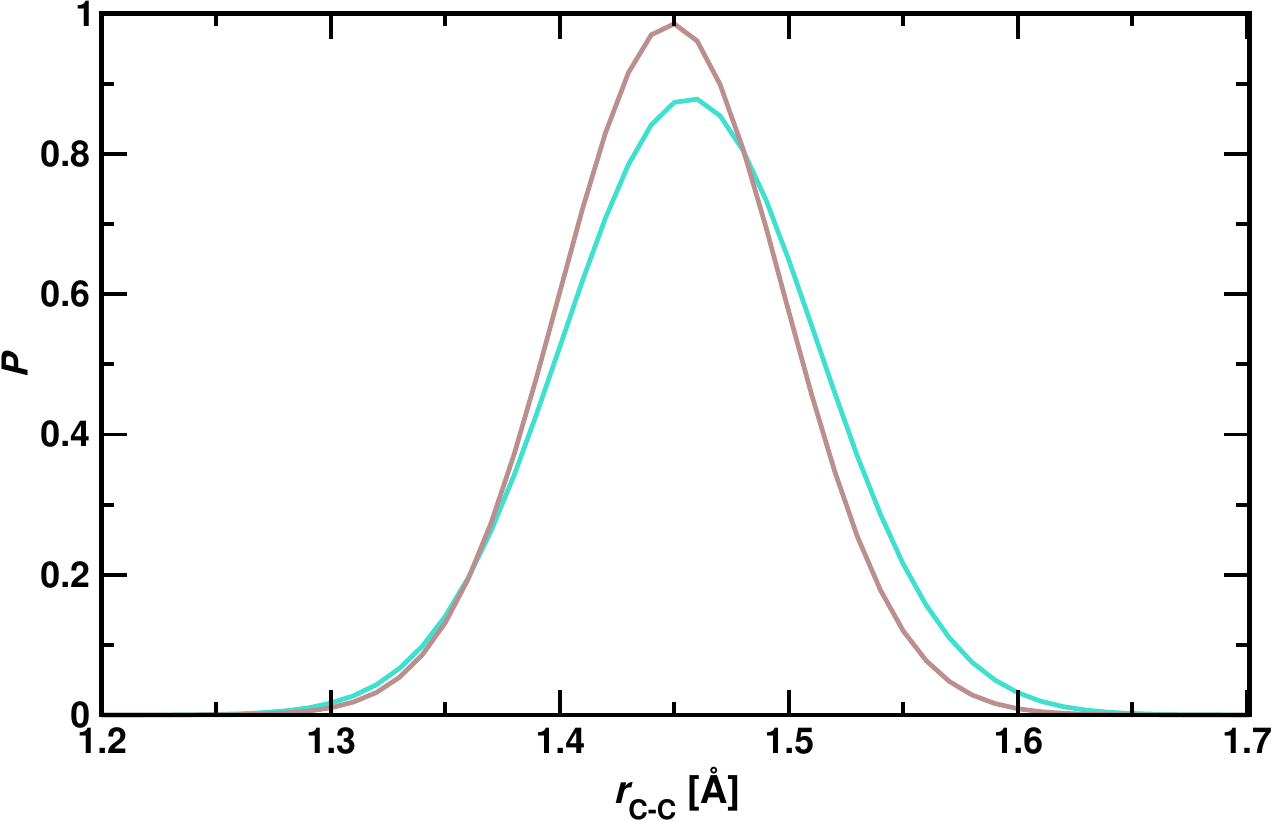}
   \caption{C-C bond distance distributions for the H$_2$C-CHO product
     for trajectories with low-(cyan) and high-(brown) TKER from
     analyzing 140 trajectories using the PhysNet PES with $D_e = 22$
     kcal/mol and CH excitation at 5988 cm$^{-1}$, see Figure
     2.}
    \label{sifig:cc}
\end{figure}

\begin{figure}
    \centering \includegraphics[scale=0.55]{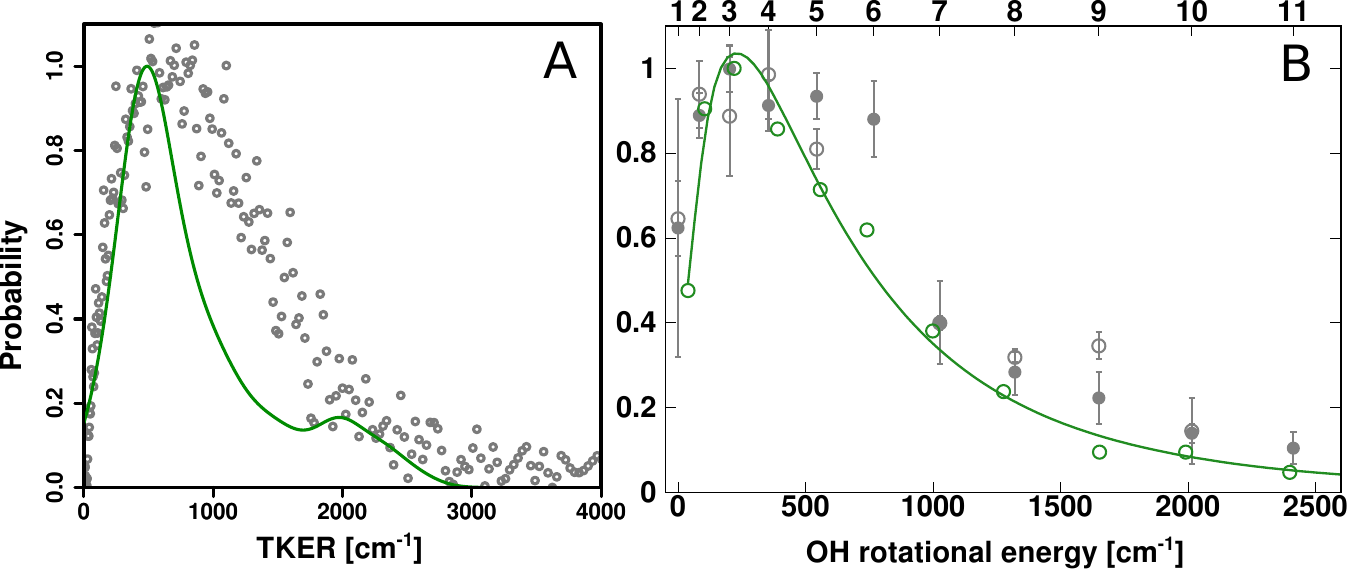}
    \caption{TKER (Panel A) and rotational state distribution (Panel
      B) for direct OH with CH-excitation at 5988 cm$^{-1}$ using
      PhysNet PES with $D_e = 28$ kcal/mol formed compared with
      experiments.}
    \label{sifig:de28}
\end{figure}

\bibliography{refs.clean}